\documentclass{emulateapj}
\usepackage{psfig}
\usepackage{graphicx}
\usepackage{color}
\usepackage{amsmath}
\usepackage{url}

\newcommand{\kms}{\mbox{km s$^{-1}~$}} 
\newcommand{\kmse}{\mbox{km s$^{-1}$}} 
\newcommand{\ms}{\mbox{m s$^{-1}~$}}

\newcommand{\vhelio}{$V_{\rm HELIO}~$}

\newcommand{\dgr}{$^{\circ}~$}

\newcommand{\teff}{$T_{eff}$~}
\newcommand{\teffe}{$T_{eff}$}
\newcommand{\logg}{$\log{g}$~}

\begin{document}

\title{The Data Reduction Pipeline for the \\
Apache Point Observatory Galactic Evolution Experiment}

\shorttitle{APOGEE Reduction Pipeline}
\shortauthors{NIDEVER ET AL.}

\author{David L. Nidever\altaffilmark{1,2},
Jon A. Holtzman\altaffilmark{3},
Carlos Allende Prieto\altaffilmark{4,5}, 
Stephane Beland\altaffilmark{6},
Chad Bender\altaffilmark{7,8},
Dmitry Bizyaev\altaffilmark{9},
Adam Burton\altaffilmark{2},
Rohit Desphande\altaffilmark{7,8},
Scott W. Fleming\altaffilmark{10,11},
Ana E. Garc\'{\i}a P\'erez\altaffilmark{2,5}, 
Fred R. Hearty\altaffilmark{2,7},
Steven R. Majewski\altaffilmark{2}, 
Szabolcs M\'esz\'aros\altaffilmark{5,12,13},
Demitri Muna\altaffilmark{14},
Duy Nguyen\altaffilmark{15},
Ricardo P. Schiavon\altaffilmark{16},
Matthew Shetrone\altaffilmark{18},
Michael F. Skrutskie\altaffilmark{2},
Jennifer S. Sobeck\altaffilmark{2},
John C. Wilson\altaffilmark{2}}


\altaffiltext{1}{Department of Astronomy, University of Michigan, Ann Arbor, MI 48109, USA
  (dnidever@umich.edu)}
\altaffiltext{2}{Department of Astronomy, University of Virginia, Charlottesville, VA 22904-4325, USA}
\altaffiltext{3}{New Mexico State University, Las Cruces, NM 88003, USA}
\altaffiltext{4}{Instituto de Astrof\'{\i}sica de Canarias, Via L\'actea s/n, 38205 La Laguna, Tenerife, Spain}
\altaffiltext{5}{Universidad de La Laguna, Departamento de Astrof\'{\i}sica, 38206 La Laguna, Tenerife, Spain}
\altaffiltext{6}{Laboratory for Atmospheric and Space Sciences, University of Colorado at Boulder, Boulder, CO, USA}
\altaffiltext{7}{Department of Astronomy and Astrophysics, The Pennsylvania State University, University Park, PA 16802, USA}
\altaffiltext{8}{Center for Exoplanets \& Habitable Worlds, The Pennsylvania State University,
  525 Davey Lab, University Park, PA-16802}
\altaffiltext{9}{Apache Point Observatory and New Mexico State
  University, P.O. Box 59, Sunspot, NM, 88349-0059, USA}
\altaffiltext{10}{Computer Sciences Corporation, 3700 San Martin Dr, Baltimore, MD 21218, USA}
\altaffiltext{11}{Space Telescope Science Institute, 3700 San Martin Dr, Baltimore, MD 21218, USA}
\altaffiltext{12}{Department of Astronomy, Indiana University, Bloomington, IN 47405, USA}
\altaffiltext{13}{ELTE Gothard Astrophysical Observatory, H-9704 Szombathely, Szent Imre herceg st. 112, Hungary}
\altaffiltext{14}{Department of Astronomy and the Center for Cosmology and
Astro-Particle Physics, The Ohio State University, Columbus, OH 43210, USA}
\altaffiltext{15}{Department of Astronomy \& Astrophysics, University of Toronto, Toronto, Ontario, Canada M5S 3H4}
\altaffiltext{16}{Gemini Observatory, 670 N. A'Ohoku Place, Hilo, HI 96720, USA}
\altaffiltext{17}{Astrophysics Research Institute, IC2, Liverpool Science Park, Liverpool John Moores
University, 146 Brownlow Hill, Liverpool, L3 5RF, UK}
\altaffiltext{18}{University of Texas at Austin, McDonald Observatory, Fort Davis, TX 79734, USA}

\begin{abstract}
The Apache Point Observatory Galactic Evolution Experiment (APOGEE),
part of the Sloan Digital Sky Survey III, explores the stellar
populations of the Milky Way using the Sloan 2.5-m telescope linked to
a high resolution (R$\sim$22,500), near-infrared (1.51-1.70 $\mu$m)
spectrograph with 300 optical fibers.  For over 150,000 predominantly
red giant branch stars that APOGEE targeted across the Galactic
bulge, disks and halo, the collected high S/N ($>$100 per
half-resolution element) spectra provide accurate ($\sim$0.1 \kmse)
radial velocities, stellar atmospheric parameters, and precise
($\lesssim$0.1 dex) chemical abundances for about 15 chemical species.
Here we describe the basic APOGEE data reduction software that reduces
multiple 3D raw data cubes into calibrated, well-sampled, combined 1D spectra,
as implemented for the SDSS-III/APOGEE data releases (DR10, DR11 and DR12).  The processing of the
near-IR spectral data of APOGEE presents some challenges for reduction, including
automated sky subtraction and telluric correction over a 3\degr--diameter
field and the combination of spectrally dithered spectra.  We also discuss
areas for future improvement.
\end{abstract}


%
%

\section{Introduction}
\label{sec:intro}


The SDSS-III/APOGEE project obtained high resolution IR spectra of over 150,000
Milky Way stars during the period 2011--2014, as described in \citet{dr12} and \citet{Majewski15}.
It is based on a bench-mounted spectrograph operating in cryogenic conditions that can obtain 300 simultaneous
spectra covering wavelengths of 1.51--1.70$\mu$m for light fed to it
from the Sloan 2.5-m telescope \citep{Gunn06} via 40-m long optical fibers. The
spectra are recorded onto three separate Hawaii-2RG (H2RG) arrays, where each
array covers a different wavelength span, with small gaps in
wavelength coverage between. The instrumental design and
performance are described in detail in \citet{Wilson15}.

Given the large amount of data expected for the entire survey, development of an automated reduction
pipeline was essential.  However, reduction of APOGEE spectra is not straightforward for several reasons:
\begin{itemize}
\item In the near-IR, telluric absorption
is significant and both spatially and temporally variable; telluric features affect a significant
fraction of the APOGEE observed spectrum.
\item The sky brightness, dominated by OH lines, is temporally and spatially variable. 
\item The instrument design delivers slightly undersampled spectra at its short wavelength end. To
avoid issues with undersampling for the stellar parameter and abundance determinations, data are
taken at two different dither positions, where the entire detector assembly is shifted by $\sim$0.5
pixel between.
\item The Teledyne H2RG arrays that are in the APOGEE instrument have some performance complications; 
in particular, some regions show significant persistence, where previous exposure to light affects
the subsequent behavior of the detector.
\end{itemize}

This paper describes the state of the APOGEE data reduction pipeline as it has
been used to produce data contained in the data releases for SDSS-III APOGEE, 
DR10 (June 2013; \citealt{dr10}) and DR12 (January 2015; \citealt{dr12}). 
As described below, there are still some areas for potential improvement
in the pipeline, but the main goal of this paper is to document the methods
used to process the APOGEE database for its public releases.
Data products in and access to the data releases
are discussed in \citet{Holtzman2015}. Additional details on the stellar
parameter and abundance analysis are presented in \citet{GarciaPerez2015}, 
and calibration and validation of the parameters and abundances are discussed
in \citet{Holtzman2015} for DR12 and \citet{Meszaros2013} for DR10.\footnote{DR11 was an internal collaboration data release
but followed the same procedures.}

\begin{figure*}[t!]
\begin{center}
\includegraphics[scale=0.45]{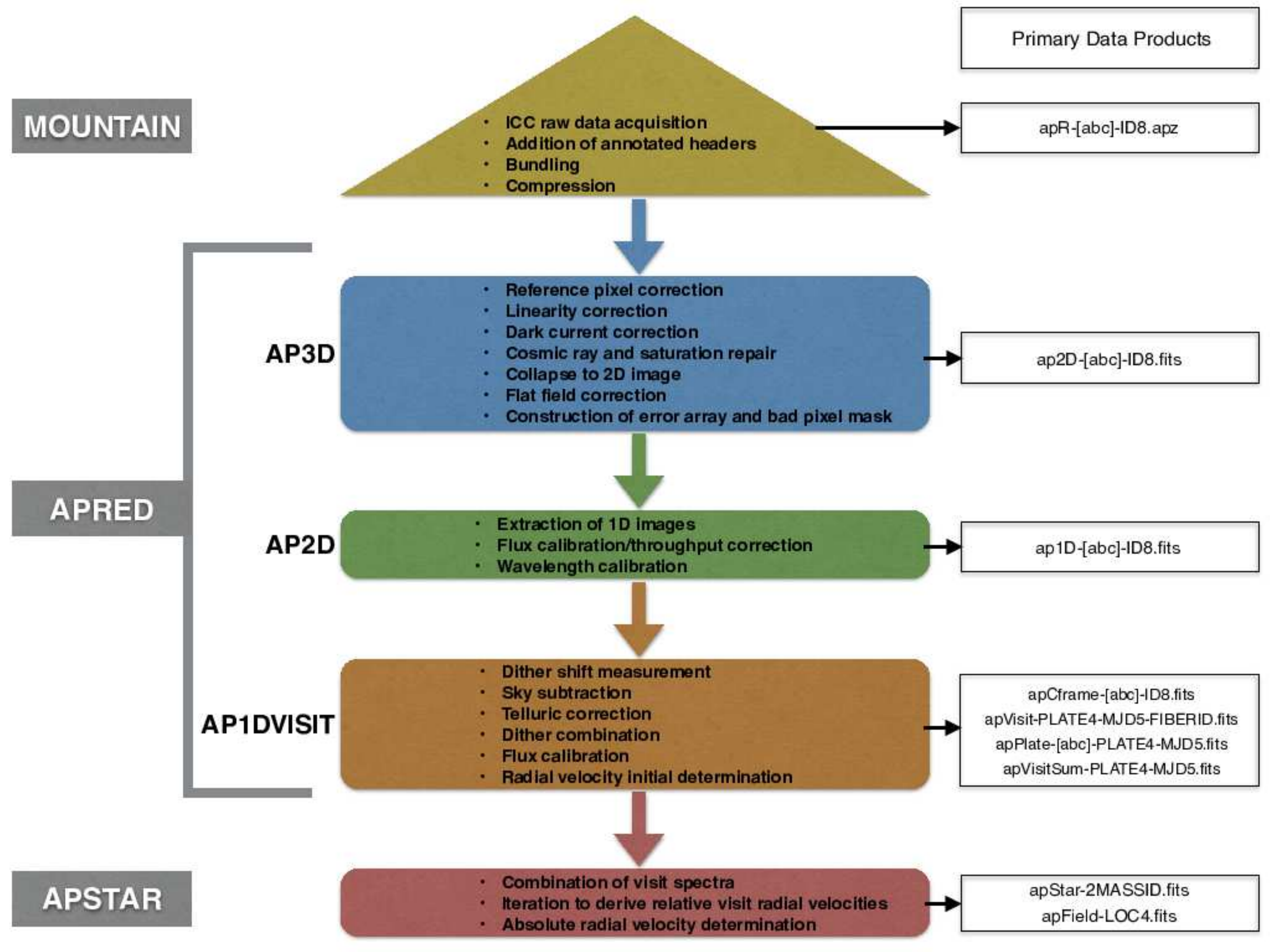}
\end{center}
\caption{Flowchart for the APOGEE data reduction pipeline listing the steps of the main stages
and the resulting data products.}
\label{fig_flowchart}
\end{figure*}

The layout of this paper is as follows.  Section \ref{sec:operations}
details the survey operations and data taking.  An overview of the
reduction pipeline is given in Section \ref{sec:overview}.  Sections
\ref{sec:ap3d}--\ref{sec:ap1dvisit} describe the three steps of APRED
that reduce observations of an individual plate on a single night:
(1) AP3D reduces the 3D data cubes to 2D images, (2) AP2D extracts the
300 spectra and determines wavelength zeropoints, and (3) AP1DVISIT
performs sky corrections, combines multiple dither exposures, and
determines initial radial velocities.  APSTAR combines spectra of
individual stars on the rest-frame and determines accurate radial
velocities (Section \ref{sec:apstar}).  Radial velocity determination
for both the AP1DVISIT and APSTAR steps are described in detail in
Section \ref{sec:rvs}.  Access to the data products is described in
Section \ref{sec:access}, and, finally, a summary is given in Section
\ref{sec:summary}.  Figure \ref{fig_flowchart} shows a flowchart of
the pipeline processing steps.

\section{Survey operations and data taking}
\label{sec:operations}

Initial commissioning data were taken starting in April 2011. The initial
data showed some issues with the internal optics and with relative focus
of the three detectors, so the instrument was opened during summer 2011
to improve these issues. It was subsequently cooled in August 2011, and
official survey-quality data began to be collected after this time. The
instrument remained cold, and in the same optical configuration over
the course of the entire survey and is therefore quite uniform.

To collect data, the 300 APOGEE optical fibers are coupled to
standard Sloan 2.5-m plug plates.  For routine first year science
observations, 230 of the fibers are placed on science targets (almost
all stars); 35 additional fibers are placed on blue stars to be
used to measure telluric absorption, and 35 fibers are placed on sky
regions without objects. Targeting is based almost entirely on the
2MASS catalog and is described in detail in \citet{Zasowski2013}.
For most survey fields, the fibers are distributed over a 3\dgr
(diameter) field of view, although for some low declination, high airmass fields,
a smaller field is used to minimize differential refraction effects.

Data are collected from the SDSS telescope using the standard SDSS telescope/instrument
interface, STUI (SDSS Telescope User Interface).  The normal mode of operation is to take individual
exposures of 500s duration. Exposures are taken at either of two dither
positions (``A'' or ``B''), where the detectors are nominally moved by $\sim$0.5 pixels
between exposures (although early survey data had a somewhat smaller shift of $\sim$0.4 due to technical issues
but this did not adversely affect the reduction).
The standard observing sequence collects 2 ABBA exposure sequences
per plate, which leads to slightly over one hour of exposure per plate
on a night. For most fields (but not all, see \citealt{Zasowski2013}), 
the target exposure time is three hours,
but these are collected over three different visits to the field, spread out in time,
to enable identification of radial velocity variation arising from stellar
binarity. Individual visits to a field are identified by a plate identification
and an MJD\footnote{Modified Julian Date (MJD) = Julian Date (JD) $-$ 2400000.5.  The MJD used by
SDSS-III is MJD$+$0.3 days so that the ``day'' increments in the afternoon at Apache Point Observatory.}.
Individual visits to an object are identified by a plate,
MJD, and fiber number; a given star will not generally be observed
in the same fiber in subsequent visits, since plates are typically replugged
between visits.



The data collection system continually reads the detectors non-destructively
as the charge is being accumulated, at slightly more than 10 seconds
between reads, so the 500s exposures are composed of a series of 47 readouts. Since
these ``up-the-ramp'' readouts are accessible as the exposure is
proceeding, it is possible to analyze count rates as the exposure
is accumulating.  This is done by ``quicklook'' software that
communicates this information to the SDSS observers. To date, this
has largely been used for informational monitoring, and total
exposure times have been fixed to 500s. Under poor conditions, a
third ABBA sequence is sometimes obtained.  During the final visit (usually
the third) to a field, only one ABBA sequence is taken if it is
likely to be sufficient to reach the total desired accumulated $S/N$.


After exposures are finished, a quick reduction is done to provide observers with
some roughly reduced data (i.e., extracted 1D spectra) to inspect.
The quick reduction software also takes the files with the individual readouts, 
bundles them into three data cubes (one for each detector), and
compresses them (see next section).
Additionally, information about the exposure and quick-reduced
spectra are inserted into a database running on the mountain.  This
database is used to monitor progress of the observations on each
field, and is the basis for the autoscheduler, which determines the
plan for plugging and observing of new fields.
A web application allows for a graphical interface to this database.

Calibration data are obtained by coupling the fiber bundle 
from the instrument to a fiber bundle that leads to an integrating 
sphere with calibration sources. Three calibration sources are
available: a continuum source, and Thorium-Argon-Neon (ThArNe) and Uraninum-Neon (UNe) hollow-cathode lamps.
In addition, several IR LEDs were installed on a cold
shutter mechanism that was installed in summer 2011. These are
located downstream of the internal slithead, and provide
roughly uniform illumination of the detectors that can be
used to determine pixel-to-pixel sensitivity variations; we
refer to such frames as internal flats.

Some calibration data are taken on a daily basis, mostly for instrument
performance monitoring. At the end of the afternoon, a few test frames
of a continuum source and some line lamp exposures are taken for quick
inspection by the observers to confirm routine instrument performance.
At the end of each night, a more complete set of calibration data
are obtained; this set includes several long dark frames, lamp exposures (ThArNe and
UNe at both dither positions), and several internal flat fields.
More extended sequences of calibration data were taken near the beginning
of the survey and were repeated periodically throughout the
duration of the survey.

In addition, on each observing night, 4 exposures (1 $\times$ ABBA) are
taken of a random pointing on the sky so that all fibers are 
illuminated by sky (predominantly OH lines). These frames are
used to characterize and monitor the image quality and the
related spectral line spread function (LSF).




%
%


\subsection{Data volume and compression}
\label{subsec:compression}

The nightly volume of data collected is significant. Each standard exposure has 47 readouts of three
2048x2048 chips, and the instrument computer also collects an additional 2048x2048 array that
provides bias information. This leads to roughly 1.5 GB per exposure.  For multiple exposures
and multiple plates, plus associated calibration data, this leads to of order $\sim$100 GB
of data per full night of APOGEE observing.

All of the raw up-the-ramp APOGEE data are kept and transferred daily off the mountain to the SAS.
To speed up the data transfer off the mountain and reduce overall disk space the raw data are
compressed using a custom designed algorithm. This algorithm takes advantage of the fact that
successive reads of the arrays are very similar; as a result, the sequence of difference
images has relatively smaller dynamic range and can be compressed efficiently.
Three steps are used to compress the up-the-ramp data cubes:
\begin{enumerate}
\item The detector reads are converted into difference images resulting in N$_{\rm reads}$ integer
images (N$_{\rm reads}$-1 difference images and the first read). 
\item The average difference image
(rounded to integers) is computed and subtracted from the difference images resulting
in N$_{\rm reads}$+1 integer images (N$_{\rm reads}$-1 ``residual'' images, one average difference image, and the
first read).  The N$_{\rm reads}$+1 integer images are written to a multi-exension FITS file.
\item The FITS file is compressed using the FPACK\footnote{\url{http://heasarc.nasa.gov/fitsio/fpack/}}
routines \citep{Pence10} and the lossless Rice compression algorithm.  
\end{enumerate}
These custom APOGEE
compressed files (a separate one for each of the three arrays) are saved to disk with
``.apz'' extensions and are on average compressed by a factor of $\sim$2.  The theoretical best compression
rate of data of our noise-level ($\sim$20) and bits per pixel (16) is $\sim$2.6 \citep{Pence09}, although
in practice the best algorithms will obtains compressions of $\sim$2.2.  Therefore, our custom compression
algorithm is about as good as can be expected for our data.

Compression and uncompression algorithms are implemented in the APZIP and APUNZIP custom IDL procedures.

\section{Pipeline Overview}
\label{sec:overview}

Data reduction is run off-site on the raw, compressed data downloaded
from the SAS. There are two main stages for basic data reduction:
\begin{enumerate}
 \item APRED reduces observations of an individual plate
on an individual night in three steps:
  \begin{enumerate}
  \item AP3D reduces the data cubes to 2D images, applying detector
  calibration products in the process. A separate error image and
  a mask of bad pixels are also calculated.
  \item AP2D extracts the spectra from each 2D image to produce 300
  well-sampled 1D spectra and corrects them for throughput variations.  AP2D also  
  performs wavelength calibration using exposure-specific wavelength zeropoints determined from the positions of sky lines
  and wavelength solutions previously derived from emission-line lamp exposures.
  Error images and masks are also produced.
  \item AP1DVISIT measures accurate dither shifts between
exposures in a visit, corrects individual exposures for sky emission
and absorption, and combines multiple dithered exposures in
a visit to produce 300 1D spectra.  The code then
determines an initial radial velocity estimate for each stellar object using
a best-matching stellar template. Outputs include
sky-corrected spectra, as well as pixel-by-pixel errors, mask information, and wavelength array, and
the correction spectra used for the sky correction. In addition,
information about the observed dithering and dither combination
are output.
  \end{enumerate}
\item APSTAR is run after multiple individual visits 
have been obtained. It resamples the spectra onto a fixed wavelength
grid (constant dispersion in $\log(\lambda)$), correcting
each visit for the visit-specific radial velocity, and coadds
the spectra. Using the combined spectra, the code derives relative radial
velocities via cross-correlation of each individual spectrum
with the combined spectrum, and puts them on an absolute scale
by cross-correlating the combined spectrum against a best-matching
template spectrum.  As a check, individual visit RVs are rederived
by cross-correlating each visit spectrum against the common template
that best matches the combined spectrum.
\end{enumerate}

A flowchart of these processing steps and the resulting data products are shown in
Figure \ref{fig_flowchart}.

A third pipeline stage determines stellar parameters and chemical
abundances in the APOGEE Stellar Parameters and Chemical Abundances
pipeline (ASPCAP), as described in \citet{GarciaPerez2015}.

The software used for the APOGEE data reduction pipeline is almost exclusively implemented 
in the Interactive Data Language (IDL).\footnote{A product of Exelis Visual Information Solutions
formerly ITT Visual Information Systems and Research Systems, Inc.} The code is archived and managed through use of the SDSS-III
software repository using the software management package Subversion (SVN). While
it was not designed for general public usage, the DR12 version of the code is publicly available
\url{http://www.sdss3.org/svn/repo/apogee/apogeereduce/}.


\section{AP3D: Reduction of data cube to 2D image}
\label{sec:ap3d}

During an exposure the three APOGEE arrays are read out non-destructively every $\sim$10.6 seconds in
sample-up-the-ramp (SUTR) mode. In addition to providing the opportunity for
inspection of data as it is being accumulated, the SUTR can be used advantageously to: 
\begin{itemize}
\item reduce the read noise by using multiple measurements of the electrons as they are accumulated \citep{Rauscher07};
\item detect and correct cosmic rays;
\item potentially correct saturated pixels if there are
enough ($\gtrsim$3) unsaturated reads to measure the flux rate, and the flux rate is assumed constant in time.
\end{itemize}

As discussed above, the separate APOGEE raw readouts are stored in a data cube (one per
array) and subsequently compressed.  In AP3D, these data cubes
are ``collapsed'' into 2D images.  Basic calibration is also done at this stage, leading
to these main steps:
\begin{enumerate}
\item reference pixel correction, 
\item linearity correction (currently not implemented),
\item dark subtraction, 
\item cosmic ray detection and repair; saturated pixel correction (currently not implemented)
\item collapse to 2D image, 
\item flat fielding,  
\item construction of error array and bad pixel mask.
\end{enumerate}

These are all described in more detail below.


\subsection{Detector electronics and reference pixel correction}
\label{subsec:refpixels}

Each of the three Teledyne Hawaii 2RG detectors (each with $2048 \times 2048$ 18$\mu$m pixels) are
read in parallel through 4 different channels per chip, with each ``quadrant'' being $512 \times 2048$ in size.

The voltage bias for the Hawaii 2RG arrays can drift slowly over time, but 
reference pixels have been implement to correct for this effect.  
There are two types of reference pixels: (1) a perimeter
of 4 pixels around each array (``embedded'' reference pixels) that are not ``active''
but are read out the same way as the rest of the array (via 4 output channels per array).
(2) A single reference pixel for each array that is read out with its own readout port, and
is called the ``reference output''.  This output channel is read out in parallel and at the same
rate as the other four output channels (leading to five altogether) producing a separate $512 \times 2048$ image. 
In the raw data cubes this extra $512 \times 2048$ ``reference'' image is attached to the end of the
regular read image to produce a $2560 \times 2048$ array.  The reference image is useful to correct for electronic
``ghosts'' that are created when very high counts from a single output affect the other three.

Both types of reference pixels are used in the APOGEE reduction.  First, the $512 \times 2048$
reference arrays are subtracted from each quadrant.  Second, vertical ramps are created
for each quadrant using means of the bottom/top reference pixels and then subtracted.
Finally, horizontal ramps are created for the entire array using 50-pixel smoothed values
from the left/right reference pixels and subtracted.  This is performed separately for each
readout and detector.


\subsection{Linearity}
\label{subsec:linearity}


Most infrared detectors have some degree of nonlinearity, with the sensitivity
changing slightly as charge is accumulated \citep{Kubik2014}. Linearity corrections are likely
to be less important for our spectroscopic analysis because the dynamic range of
a given spectrum, especially over the portion of any individual spectral features,
is generally relatively small, and we have no requirement for high accuracy in
relative flux between different objects.  Taking the nonlinearity coefficients from \citet{Kubik2014},
the error in the relative depth of a 20\% (of the continuum) absorption line for a spectrum with a
continuum of 10,000 ADU is $\sim$0.4\%.  The large majority of our stellar spectra have continua
(in individual exposures) below this level, and, therefore, the effects of nonlinearity are minor.

We have made some initial tests for nonlinearity using internal flat field data
cubes, under the assumption that the LED light source is stable (which it 
appears to be, judging from the repeatability of light levels in
successive exposures). These data suggest that there may be some
small level of nonlinearity, but characterizing it is complicated
by the behavior of some pixels at low light levels. As a result,
we have chosen not to implement any nonlinearity correction at this
time.


Additionally, as discussed below, some regions of two of the detectors suffer
from a significant persistence effect, where the amount of charge deposited
can be affected by the previous exposure. This effect is significantly
larger than any expected non-linearities in these regions.

Consequently, although the pipeline has an implementation for a linearity 
correction, we have not applied such a correction for the DR10--DR12 data.

\begin{figure}[t!]
\begin{center}
\includegraphics[scale=0.55]{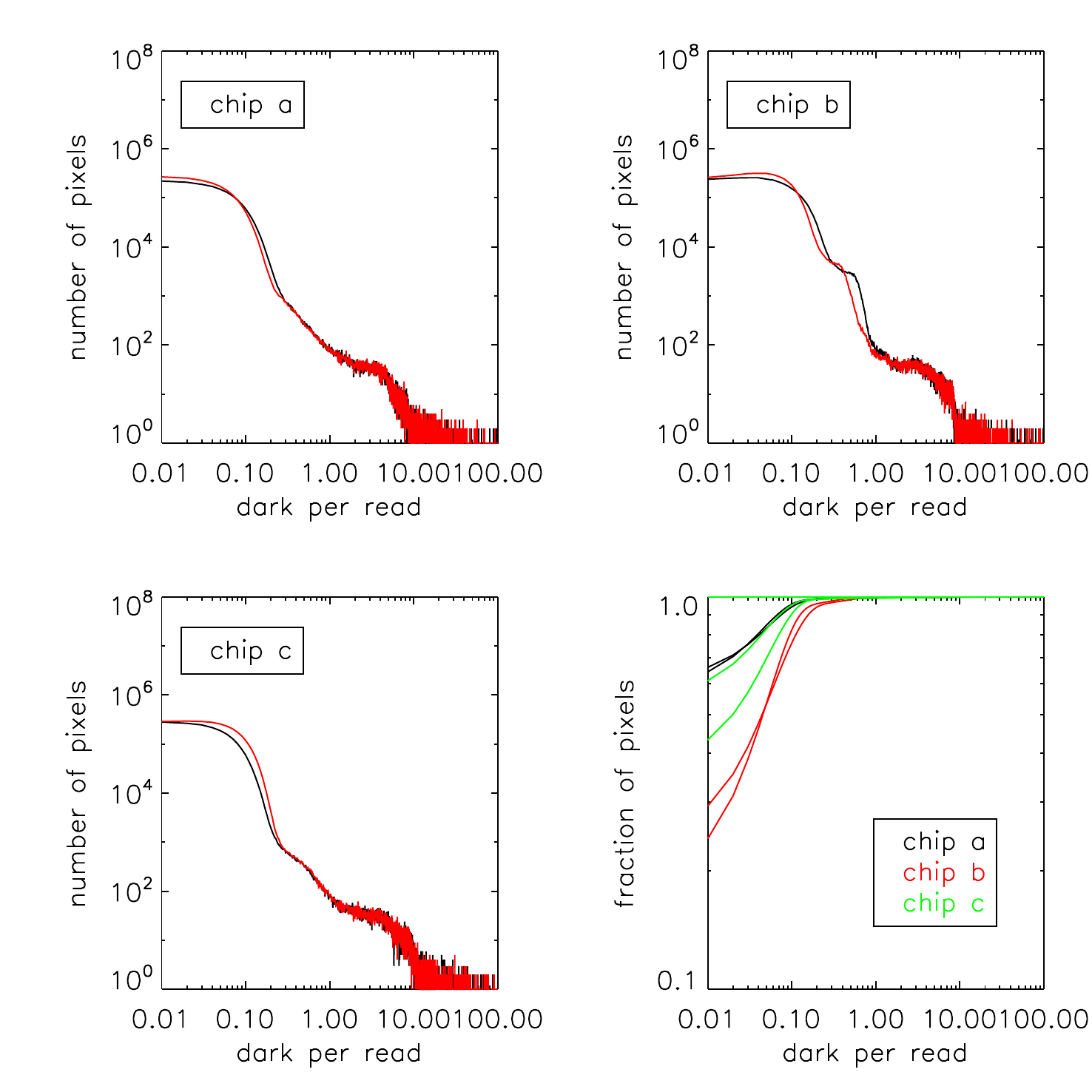}
\end{center}
\caption{Histogram of dark current rates (DN/read). Three panels show
histograms for the three chips respectively; the lower right shows the cumulative
histograms for all three. Data are shown for two different dates separated
by over two years.}
\label{fig:dark}
\end{figure}

\subsection{Dark current}
\label{subsec:darkcorr}

The dark current is derived from multiple 60 read exposures with the
internal cold shutter closed. Figure \ref{fig:dark} shows
the distribution of the dark current rate (counts per read) 
of all of the pixels. Three panels show the histogram of
dark rates for each of the three chips, whereas the lower right shows
the cumulative distribution across all pixels for all three chips.  While most pixels
have a dark rate below 0.5 counts/read, there is a tail up to high
dark rates for some pixels. In fact, the typical dark current is significantly lower
than 0.5 counts per read because, even after averaging 20--30 frames to reduce the noise,
readout noise dominates over dark current at this level. The middle array has a section
of higher dark current than the other two, and this is reflected in
the histogram.  For the pixels with very high dark current, the
effect of interpixel capacitance (IPC, see \citealt{Rauscher07}) is
clearly visible because the hot pixels produce small crosses on the
detector, with charge coupled to the adjacent pixels.

To correct for dark current, ``superdark'' frames are constructed
by taking the median of 20 long dark frames.
To allow for the possibility that dark current may
not accumulate linearly with time, the superdark is constructed
for each up-the-ramp readout, so the superdark calibration
frame is a data cube, with the appropriate slice subtracted
from the corresponding readout of each science frames.

\begin{figure}[t!]
\begin{center}
\includegraphics[angle=0,scale=0.48]{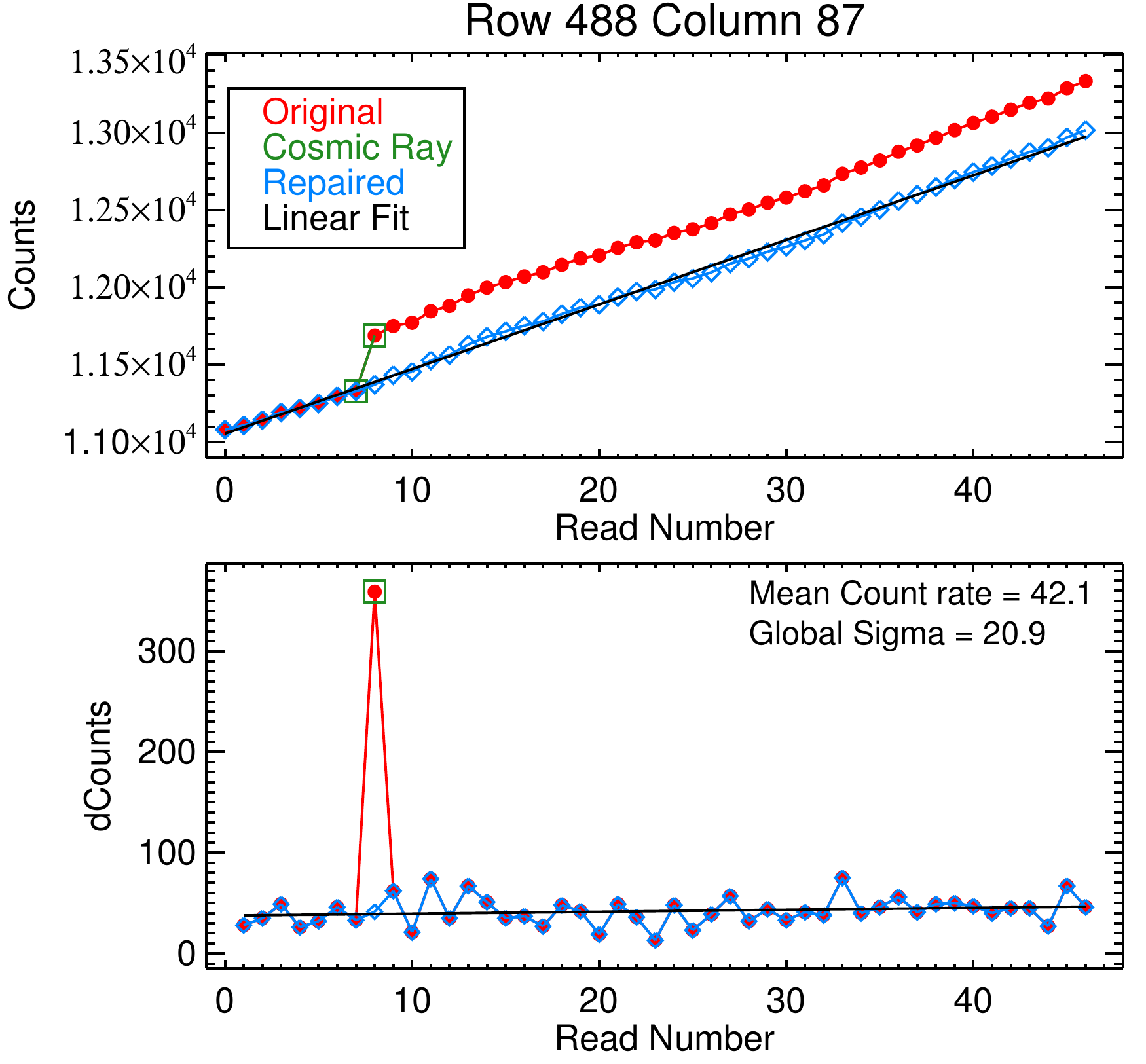}
\end{center}
\caption{Example of cosmic ray detection and ``repair''.  (Top) Counts and (bottom) count rate
  as a function of read number.  The original values (red), repaired values (blue), detected cosmic ray (green),
  and linear fit to the fixed values (black) are shown.}
\label{fig_crfix}
\end{figure}

For the current analysis, we have used a single superdark, constructed
from data taken on MJD 56118. Figure \ref{fig:dark} (top and lower-left panels) shows the histograms
from this date in black, as well as the histograms from darks taken at the
end of the SDSS-III/APOGEE survey, on MJD 56853 (in red).  Comparison of these
histograms, as well as direct comparison of the dark images, show that
the dark current is quite stable, with only a few pixels changing their dark
rate significantly.  Daily dark frames are taken as part of the normal 
calibration, and future analysis may use these to implement correction
for small changes in dark rates and the number of hot pixels.

All pixels with a dark rate exceeding 10 counts/read are marked as
bad pixels.  Neighbors of any of these are marked as bad if their dark
rates exceed 2.5 counts/read.

%



\subsection{Cosmic Ray and Saturated Pixel Correction}
\label{subsec:crcorrection}

\begin{figure}[ht!]
\begin{center}
\includegraphics[angle=0,scale=0.50]{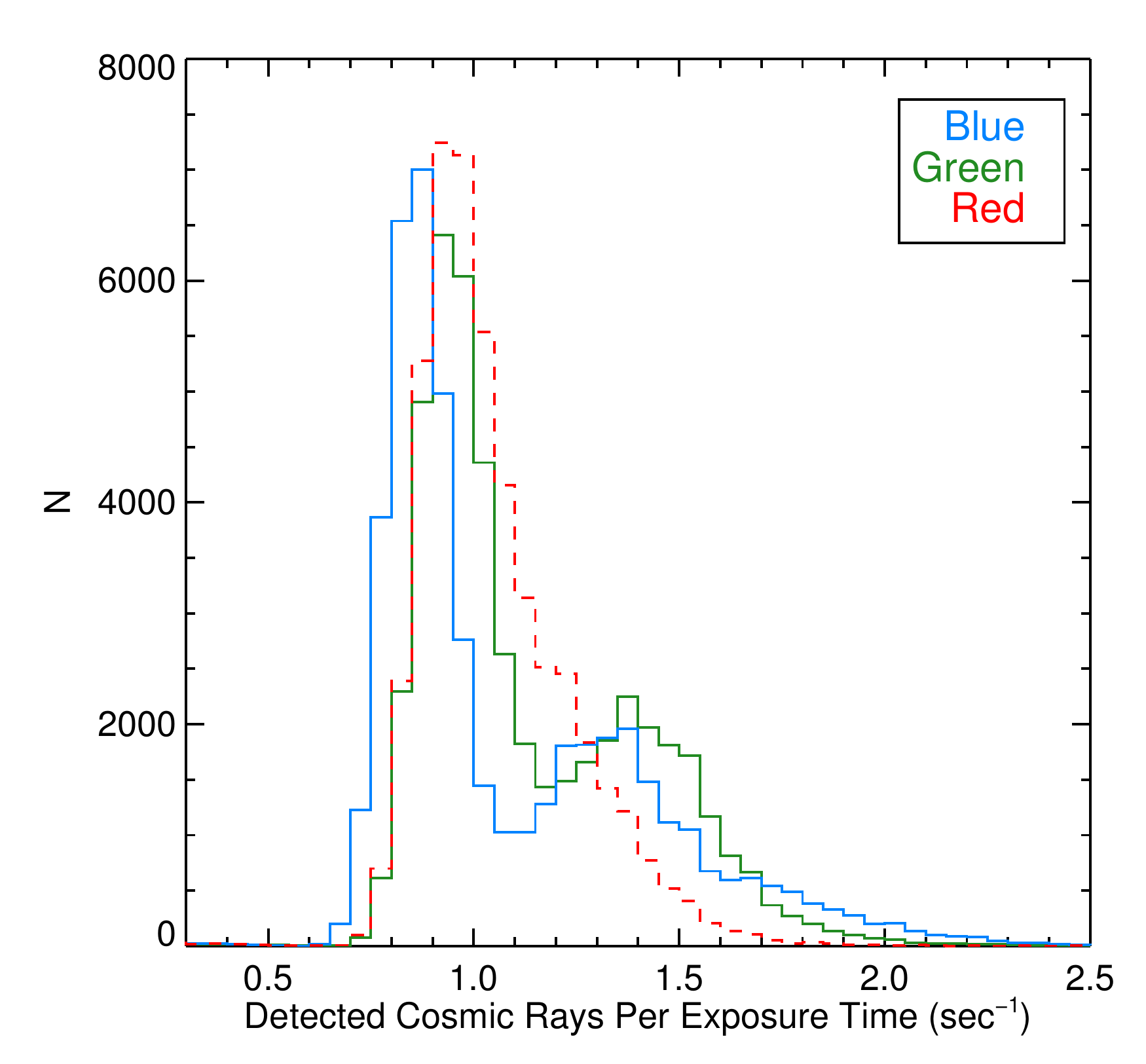}
\end{center}
\caption{Histogram of cosmic rays detected per exposure time on the three APOGEE detectors ($\sim$1359 mm$^2$).  The median
values are 0.92/1.00/0.99 (blue/green/red).  The images with shorter exposure
times have slightly higher rates likely because of lower Poisson noise that allows
for the detection of weaker cosmic rays.}
\label{fig_crhist}
\end{figure}

One advantage of using SUTR sampling is that cosmic rays can be detected
and removed.  Each pixel is searched for positive jumps that could correspond to cosmic
rays. First, the array of read values is turned into difference counts (between successive
reads).  Second, a median filtered version of the difference counts is created (using a median filter
of 11 reads) that can be used to remove any flux rate variations over time (e.g., from seeing
variations or clouds).  Then, a ``local'' scatter of the difference counts around the local median value is
measured using a robust standard deviation.  Any difference count values larger than
10$\times$ the local scatter above the local median (and 10$\times$ above the noise level) are
flagged as cosmic rays.  The cosmic rays are corrected by replacing their difference values with the local
median for that pixel.  Finally, the counts array is reconstructed by adding the cumulatively summed difference
counts to the first read value.  The pixels with detected cosmic rays are flagged in the mask image.
This cosmic ray detection method will miss some very weak cosmic
rays but should catch most of the large ones that are more likely to affect the data.  Figure \ref{fig_crfix}
shows an example of the cosmic ray correction process.  Figure \ref{fig_crhist} shows the distribution
of detected cosmic ray rates in the DR12 data for the three arrays.  The median detected cosmic rays
in object exposures (500s) are 430/470/470 (blue/green/red).

Another advantage of using SUTR sampling is that measurable signal is recorded 
even for pixels that end up being saturated after the full exposure time.  If as few as $\sim$3--4
reads are unsaturated, then the flux rate can be measured and used to extrapolate the counts to the
end of the exposure.  However, this extrapolation assumes that the count rate is
stable, which is not the case in sub-optimal observing conditions.  While it might be possible
to characterize the count rate variations using other pixels, the situation is complicated
because different pixels may have different variations in count rate, e.g., the rate in
spectral regions including sky emission are likely to vary in a different way from the rate in
regions of stellar signal.

The pipeline currently corrects any saturated pixels assuming a constant flux
rate, but flags such pixels as having been corrected. Because of the possibility
that this correction is a poor approximation, subsequent stages of the reduction treat
these pixels as bad.

\begin{figure*}[ht!]
\begin{center}
\includegraphics[angle=0,scale=0.58]{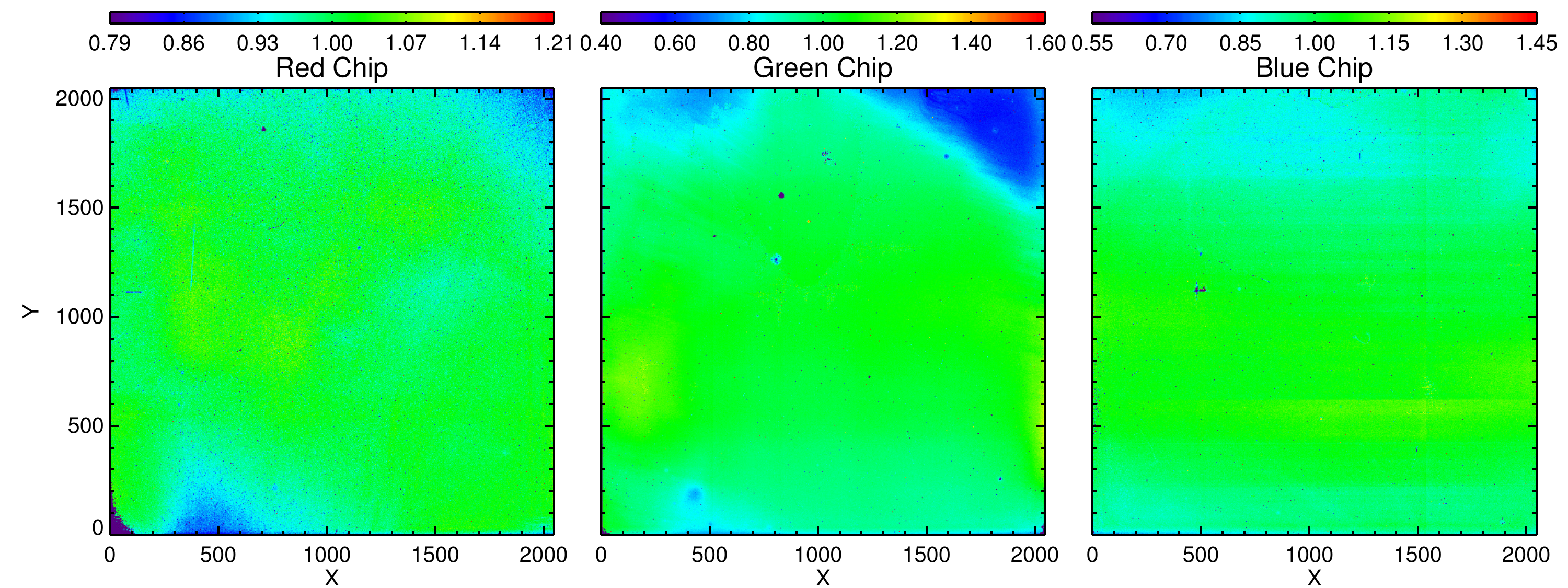}
\end{center}
\caption{The APOGEE DR12 flat field images.  Note the different colorbar ranges for each chip.
The standard deviations are 3.0, 8.2 and 7.6\% for the red, green and blue detectors.}
\label{fig_flats}
\end{figure*}

\begin{figure}[h!]
\begin{center}
\includegraphics[scale=0.58]{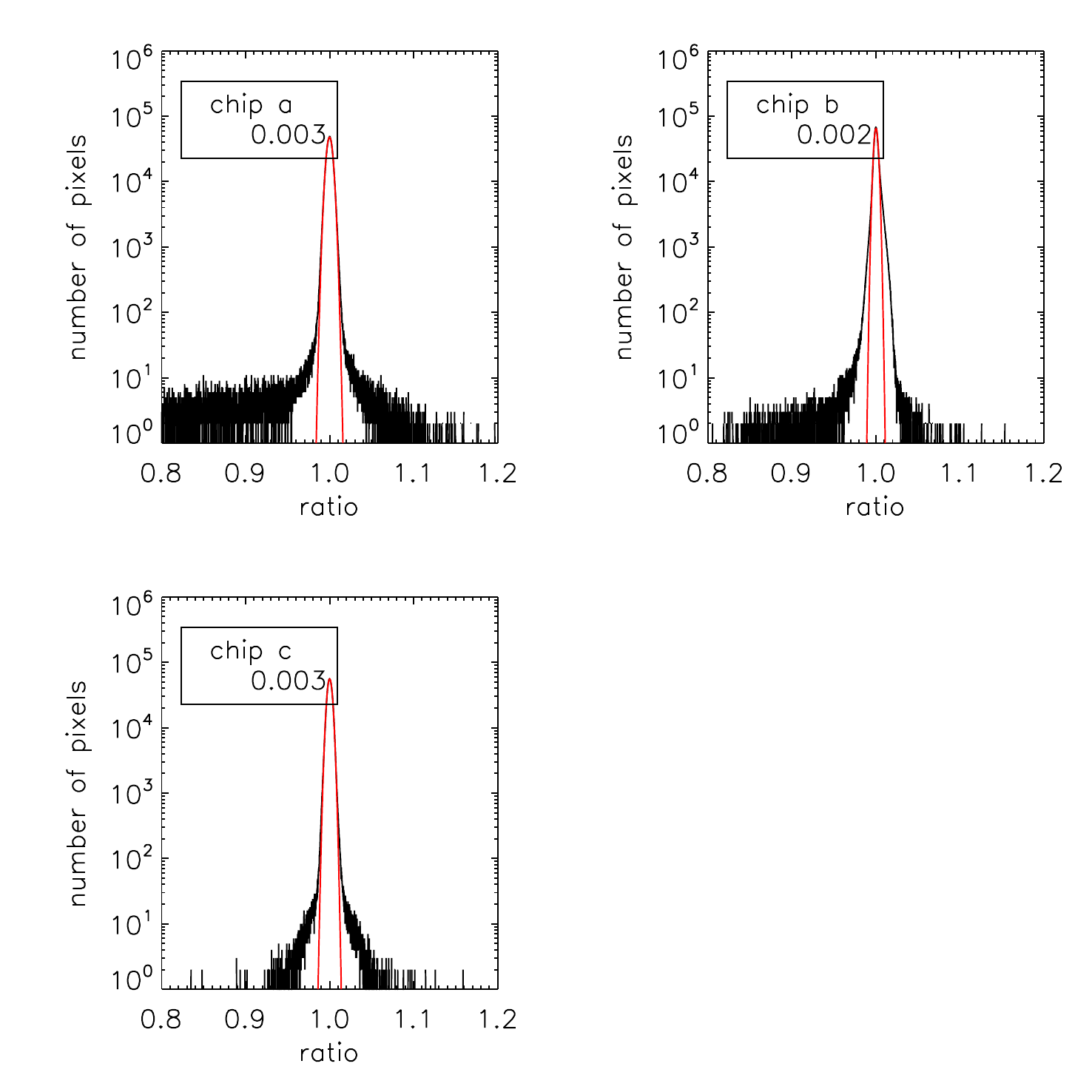}
\end{center}
\caption{Histogram of the ratio of two superflats, taken over two years
apart. The red lines show gaussian fits to the histograms, and the $\sigma$
of the fit is shown. Photon statistics alone are expected to yield 
$\sigma=0.002-0.003$.}
\label{fig:flatratio}
\end{figure}

\subsection{Collapse to 2D Image}
\label{subsec:sampling}

As is standard with non-destructive readout IR arrays, the array reset at the beginning of each
exposure introduces a pedestal value with significant ``noise'' (i.e., pixel-to-pixel variations),
so multiple non-destructive readouts are required to remove the baseline values and this ``reset noise''.
To minimize this non-negligible readout noise, 2D IR images are typically calculated from the 3D data cubes using either
Fowler sampling, where some number of readouts at the beginning and end of the exposure are averaged
and a difference image created, or using the full up-the-ramp sampling to determine a mean count
rate that, when multiplied by the exposure time, gives the total counts per pixel.  The simplest
Fowler sampling, using just one readout at the beginning
and one at the end, corresponds to correlated double sampling (CDS). A detailed
analysis of the noise using different readout methods is presented in \citet{Rauscher07}. 



Both Fowler sampling and up-the-ramp analysis are implemented in the pipeline. Up-the-ramp is used
for all of the data except for the ``dome" flats; these use simple CDS sampling because
the lamp is only turned on for a few seconds to accumulate the desired number of counts; because the
count rate is thus highly non-uniform, up-the-ramp sampling fails in this case.
In addition, the relatively large number of counts in these exposures is not significantly affected
by readout noise.

\subsection{Flat Fielding}
\label{subsec:flatfield}

Flat fields to correct for pixel to pixel sensitivity variations
are constructed using a median of ten internal flat fields and
stored in the apFlat calibration files.
Figure \ref{fig_flats} shows the observed sensitivity variations
of the three Hawaii 2RG detectors which can be quite large.
The middle (``green'') chip shows a rim of lower QE (quantum efficiency) giving the appearance of a ``thumb-print''.

The 2D images are flat fielded using a ``superflat'' constructed by the
average of 20 internal flat field frames, with large scale structure
removed; these are stored in apFlat files. The count levels in the
individual flats are such that the combined flat should have
S/N$>$ 500, or an rms of $<$0.2\%.

The flat field is very stable
in time. This is demonstrated in Figure \ref{fig:flatratio}, which shows
histograms of the ratio of a superflat constructed on MJD 56037 with those taken at the
end of the survey on MJD 56852. Given the S/N of the individual superflats,
the ratio of the two flats is expected to have $\sigma\sim 0.003$ on the
basis of photon statistics only. Figure \ref{fig:flatratio} shows that
the distribution of values in the ratio closely matches a gaussian with
width comparable to this, demonstrating that the flats are likely to be
extremely stable. We note some deviations in the 'b' chip; upon inspection,
these arise from the rim noted above, which is likely more related to 
persistence effects in this detector (see Section \ref{subsec:persistence})
than to temporal QE variations.

Any pixels in the superflat that have sensitivity less than 75\% of pixels
in their vicinity are marked as bad pixels, and saved in the bad pixel
masks.

\subsection{Gain, Noise Model and Bad Pixel Mask}
\label{subsec:error}

The detectors are read in parallel through 4 different
channels per chip.  The inverse gain ($e^-$ / DN) for
each channel in each chip is derived from pairs of internal
flat fields using the measured variance. In 50 different
intensity bins with different mean intensities ($I$), the 
variance in the difference image is measured and the inverse gain calculated 
from $2 I / \sigma^2$.  We use a gain of 1.9 for all quadrants.  This gain
calculation will be updated in the future.


A noise model is constructed by combining in quadrature the Poisson noise of the 2D image,
the Poisson noise of the dark image, and the sampling read noise (using the equations from
\citealt{Rauscher07}).  This noise image is placed in the ``error array''.


Bad pixel mask files for each detector are created using pixels marked
as bad during the construction of the superdark and superflat frames. The
bad pixel masks are saved as bitmasks to preserve the reason that any given
pixel was marked as bad, and these values are propagated via the bitmasks
to the reduced data frames.

The bad pixel mask from the apBPM calibration product is used to mask out (set to NAN) bad pixels
from each data cube and flagged in the mask image.  These pixels are not used in any of the subsequent
analysis.  

Pixels in the region of the Littrow ghost\footnote{The Littrow ghost is formed by dispersed light being reflected from the detector
arrays, reflectively recombined by the VPH grating, and finally reimaged onto the detectors \citep{Burgh2007,Wilson15}.}
($\sim$1.604 $\mu$m, but it varies with fiber), are flagged in the mask but are not marked as bad in the image since the effect
can be negligible depending on the spectrum.



\begin{figure}[t]
\begin{center}
\includegraphics[angle=0,scale=0.38]{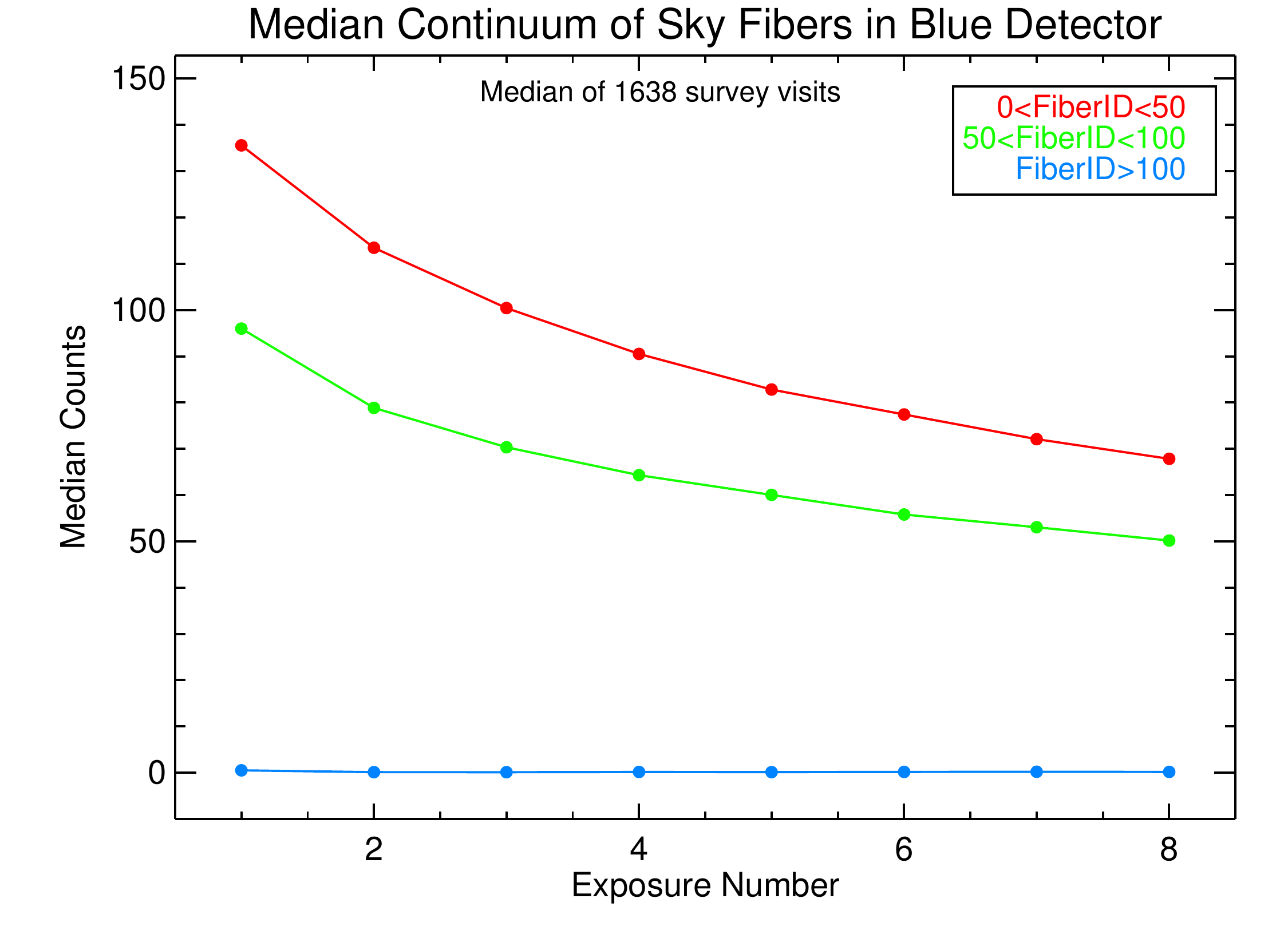}
\end{center}
\caption{The median continuum of sky fibers on the blue detector versus exposure
number in a visit (median across 1638 survey visits) for moderately high persistence region (green),
highest persistence region (red), and normal persistence region (blue). The magnitude and a rate of (temporal) decline of
persistence (mostly due to the preceding dome flat) can be seen clearly.  By the end
of a visit the persistence declines to 50--70 counts per exposure.}
\label{fig_persistence_skyfibers}
\end{figure}

\subsection{Persistence}
\label{subsec:persistence}


Teledyne H2RG arrays, like many IR detectors, exhibit persistence behavior in which a latent image of a
previous exposure appears in subsequent images but at a fraction of the original source flux
or stimulus \citep{Smith08a}.  In H2RGs, the persistence generally decays exponentially with time
with the resulting persistence representing a small percentage of the stimulus, although it can
sometimes take a long time to be released.

Normal persistence is not significant for most APOGEE exposures.
The total persistence accumulated (in a dark exposure) 1800s after a ``stimulus'' exposure is only $\sim$60 counts
even for stimulus well depths of 25,000 counts.  However, some regions of the APOGEE detectors ---
the top $\sim$1/3 of the ``blue'' array and around the perimeter of the ``green'' detector --- have
unusually high levels of persistence (which we sometimes refer to as ``superpersistence'').  In these
regions the total accumulated persistence 1800s after an exposure is $\sim$10--20\% of the stimulus counts.

\begin{figure}[t]
\begin{center}
\includegraphics[angle=0,scale=0.38]{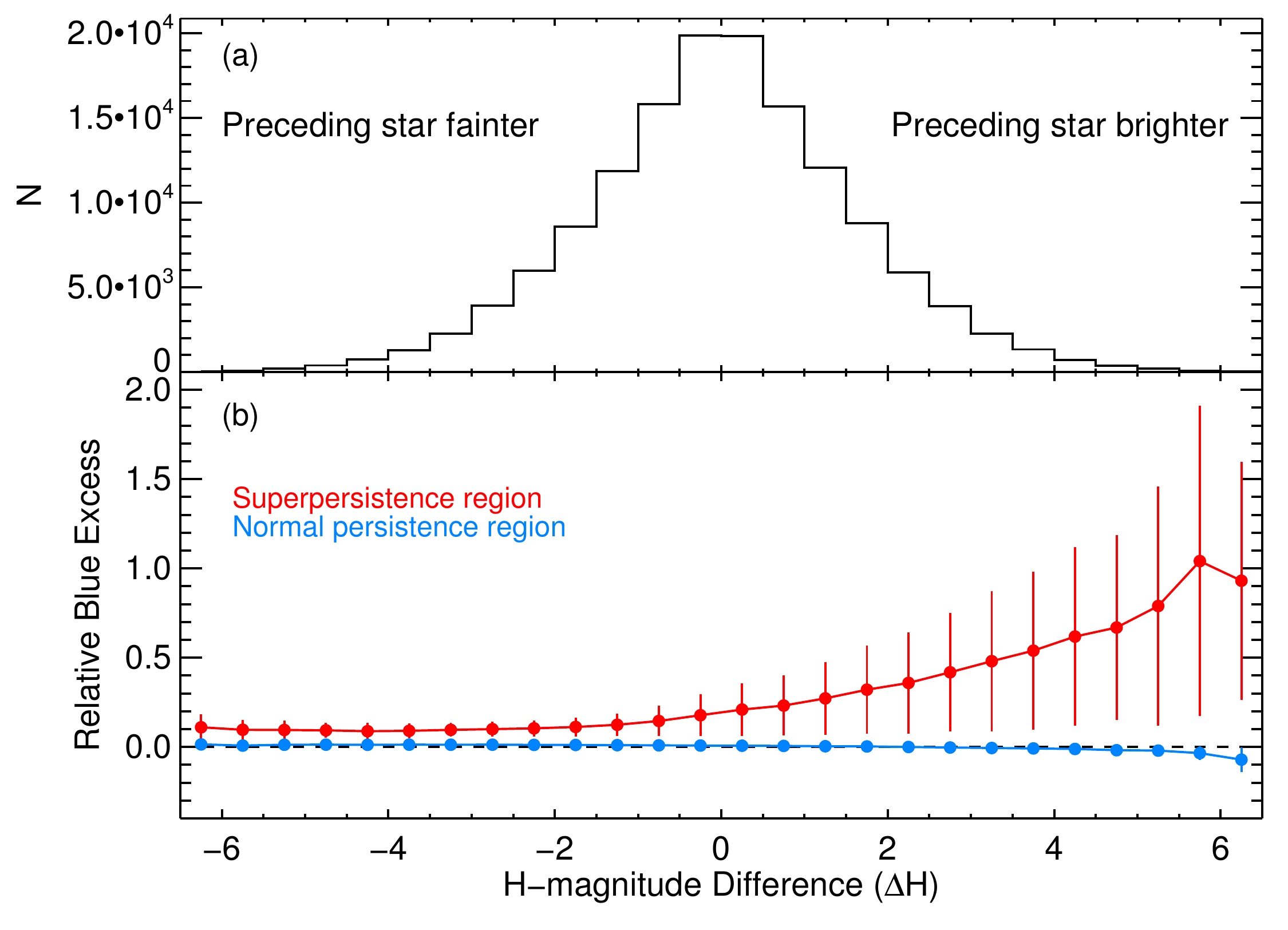}
\end{center}
\caption{(a) The distribution of magnitude difference ($\Delta$$H$) between stars in the blue superpersistence
region and the star in the preceding plate visit (in the same fiber) where a
positive difference indicates that the preceding star was brighter. (b) The relative
blue excess of the stellar spectra, (blue$_{obs}$-blue$_{mod}$)/blue$_{mod}$, where
blue$_{mod}$ is the expected (model) blue flux and is the product of the observed green
flux and an empirical relation (found from the normal persistence region) of the
blue/green flux ratio as a function of $J$-$K_{\rm s}$ color (1.092$-$0.0737[$J$-$K_{\rm s}$]).
The relative blue excess signifies the level of persistence relative to the
stellar spectrum.  The superpersistence region is shown in red and the normal persistence region in
blue.  The line indicates the median of all stars at the same magnitude difference and the error
bars the robust standard deviation.}
\label{fig_persistence_magdiff}
\end{figure}

APOGEE spectra can be contaminated in the high persistence regions in two ways: (1) by stars in the 
same fiber from preceding plate visits (most significant when the preceding star is bright),
and (2) by calibration exposures, such as the ``dome'' flats taken before each plate visit
for throughput calculations.  While the persistence from the former have spectral features that will
corrupt a stellar spectrum, the latter are generally featureless and act as a veiling component.
Fortunately, APOGEE uses a fiber management scheme in which fibers are designated for ``bright'', ``medium'' and ``faint''
stars (although each group still spans a significant magnitude range) to help reduce cross-contamination between spectra
on the detectors (i.e., to minimize the ``spatial'' wings of a bright star contaminating the spectrum of an adjacent
faint star; see \citealt{Majewski15} for more details).
This observing scheme fiber management scheme also helps reduce the relative effect of persistence from preceding stars
because the magnitude differences are smaller.

\begin{figure}[t]
\begin{center}
\includegraphics[angle=0,scale=0.38]{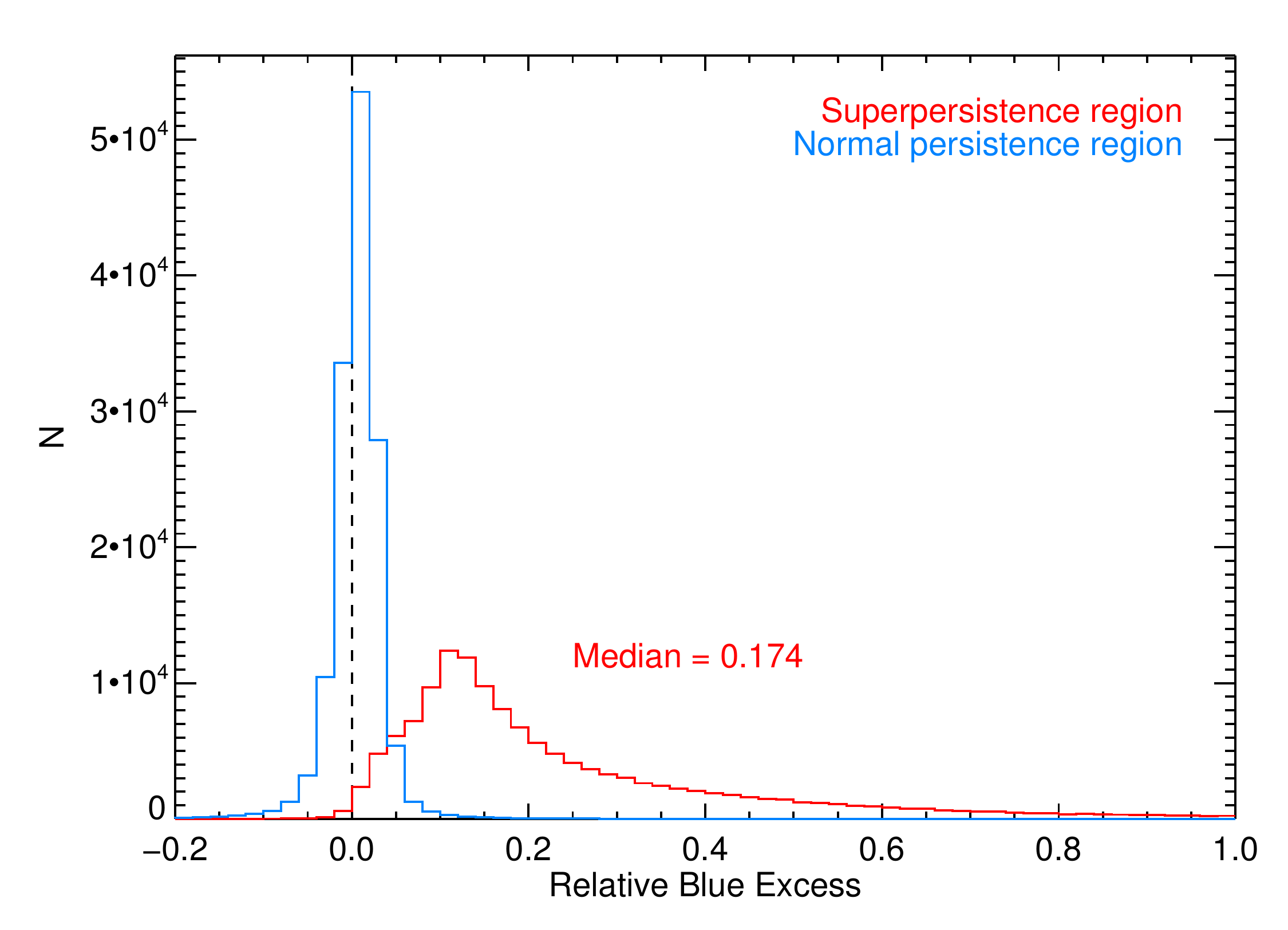}
\end{center}
\caption{The distribution of relative blue excess for the superpersistence region (red)
and a similarly sized portion of the normal persistence region (blue).  The flux ratios are higher
in the superpersistence region due to persistence from the dome flat and the preceding star in the
same fiber.  The superpersistence distribution peaks at $\sim$0.174 but has a long tail to higher
values.}
\label{fig_persistence_magdiff_relbluexhist}
\end{figure}

\begin{figure}[t]
\begin{center}
\includegraphics[angle=0,scale=0.40]{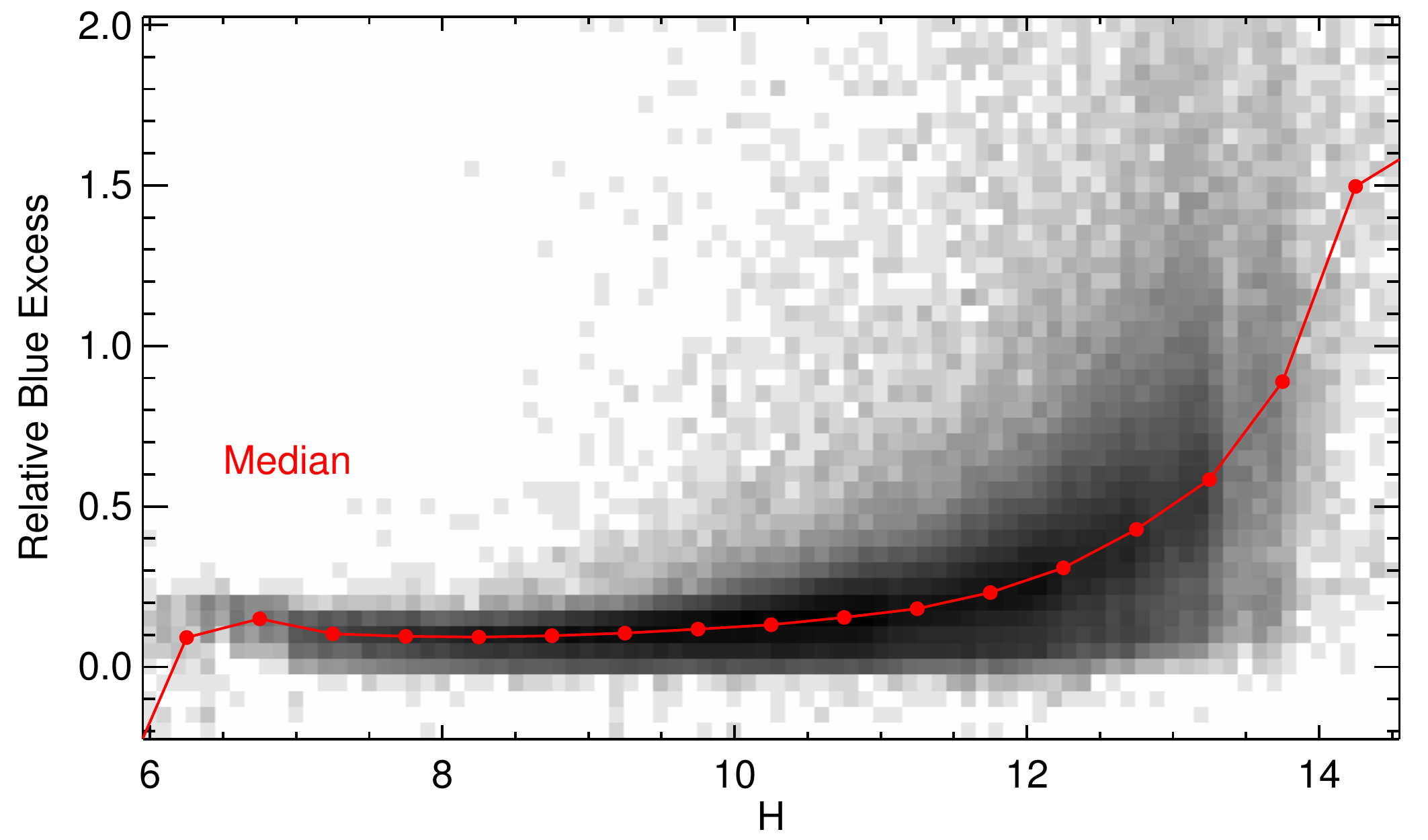}
\end{center}
\caption{The distribution of relative blue excess with $H$-magnitude for the superpersistence region.
The connected red dots show medians in 0.5 mag bins of $H$.  The persistence is most significant for
the faintest APOGEE stars.}
\label{fig_persistence_magdiff_relbluexhmag}
\end{figure}

Figure \ref{fig_persistence_skyfibers} shows the median continuum in the blue detector for sky fibers
(which should show persistence from previous exposures)
versus exposure number for many science visits.  This figure illustrates the level (and temporal behavior) of the
persistence (for which the accumulated charge is dominated by the dome flat immediately preceding a science exposure),
which mostly affects the fainter stars.  While featureless
(gray) persistence (as expected from the dome flat) will not significantly impact the measured radial velocities,
it will affect the
relative depth of the spectral absorption lines and therefore the derived abundances.  The effects of
persistence on DR12 abundances are described in \citet{Holtzman2015}.
Figure \ref{fig_persistence_magdiff}a shows the distribution of magnitude difference ($\Delta$$H$=$H_2$-$H_1$)
between a star and the previous star in the same fiber (from a different plate) while the ``relative blue excess'',
a measure of persistence relative to the stellar spectrum, is shown in Figure \ref{fig_persistence_magdiff}b.
Even when the preceding star is significantly fainter there is excess flux in the blue ($\sim$10\%), likely
due to persistence from the dome flat.  As $\Delta$$H$ increases (i.e., the preceding star gets relatively
brighter) the flux ratio also increases due to the persistence from the preceding star.  The stellar and
dome flat persistence reach parity at $\Delta$$H$$\approx$$+$0.10 mag and a relative blue excess of 1
(where the persistence is as bright as the star itself) is reached at $\Delta$$H$$\approx$$+$6 mag.
The median relative blue excess in the superpersistence region is $\sim$17.4\% but with a long tail to
higher values (Figure \ref{fig_persistence_magdiff_relbluexhist}) and, as might be expected, the faintest
APOGEE stars are the most affected by the persistence (Figure \ref{fig_persistence_magdiff_relbluexhmag}).

Our initial investigation of the APOGEE superpersistence behavior indicates that the persistence from a
single stimulus (e.g., a flat exposure) is well described by a double-exponential in time (as also seen by
\citealt{Smith08b}) with the timescales being fairly constant across pixels and stimulus ($\sim$120s and
$\sim$1700s).  The exponential amplitudes are a complex function of fluence (or count rate) as well as total
exposure time, and the behavior changes once saturation is reached.  However, persistence becomes significantly
more complex once multiple stimulus exposures are taken in a row.  While this is the information that is
required to properly correct APOGEE data for persistence (i.e., the persistence due to the multiple previous
stimuli), we have not yet fully characterized this behavior.

While there is evidence that persistence can be calibrated out in some circumstances
(e.g, the $HST$ WFC3 team has developed an algorithm that removes $\sim$90\% of the persistence in WFC3/IR
images\footnote{\path{http://www.stsci.edu/hst/wfc3/ins_performance/persistence/}}),
the APOGEE reduction pipeline for DR12 contains no correction for persistence.  This is largely due to the
complexity of the problem.  Our initial investigation has shown the WFC3 prescription to be inadequate for
the APOGEE detectors.  Work is proceeding to fully characterize the APOGEE persistence behavior and to
develop an algorithm  to correct for persistence in later data releases.
For DR12, spectra in the persistence region and with jumps in the spectral continuum from green to blue
are flagged (see \citealt{Holtzman2015}) and should be treated with caution.
Of course, the red and most of the green portions of these spectra are uncontaminated by superpersistence and
can be used for science.
Looking forward, persistence is expected to be less of a concern in the SDSS-IV/APOGEE-2 survey because the
blue detector was replaced in summer 2014 with a detector exhibiting normal persistence behavior (although the
original green detector with higher persistence around the perimeter still remains).





\subsection{Output}

All of the above tasks are performed by the IDL routine AP3DPROC program in the 
reduction pipeline.  The output files are called ap2D-[abc]-ID8.fits (the abc denoting the
red/green/blue detectors, respectively) and have three data extensions, one each for the flux, errors,
and bitwise pixel mask, and each with a size of 2048$\times$2048. These files are output to the
{\tt spectro/v\#/red/MJD5/} directory.

\section{AP2D: Extraction to 1D spectra}
\label{sec:ap2d}


Once the 2D images are created, the next step is to extract the 300 fiber spectra,
correct them for fiber-to-fiber throughput variations, and wavelength calibrate
them.  This step is performed by the IDL AP2DPROC routine.

\subsection{Extraction}
\label{subsec:extraction}

The spectra are recorded roughly along rows on the three detectors.
The standard extraction method uses an empirical measurement of
the spatial profile of each trace at each column of each detector.
Use of an empirical spatial point spread function (PSF)
is possible because this PSF is quite stable over the course
of an exposure sequence (visit), because the instrument is stationary and fiber fed. Over
a longer period, the traces do move by a fraction of a pixel as the
weight of the LN$_2$ tank suspended below the cold plate varies throughout a night
due to LN2 depletion and then refill.  However, this longer period drift is tracked using
individual ``dome flat'' exposures that are taken at the end of every
exposure sequence. These exposures are taken of 
the telescope mirror covers illuminated using a continuum source, and are used both for spatial
profile construction as well as for mapping the system fiber-to-fiber
throughput variations.

The 300 spectra are separated by roughly 6-7 pixels on average. In detail,
as described by \citet{Wilson12}, the spectra are grouped in 10 blocks
of 30 fibers each, with slightly larger gaps between these groups.

To allow accurate measurements of the wings of the PSF, data
are periodically taken using the so-called ``sparse pack'' calibration
channel, in which only 50 of the 300 fibers are populated. These
provide widely separated spectra from which the PSF can be determined accurately.




\begin{figure}[t!]
\begin{center}
\includegraphics[angle=0,scale=0.42]{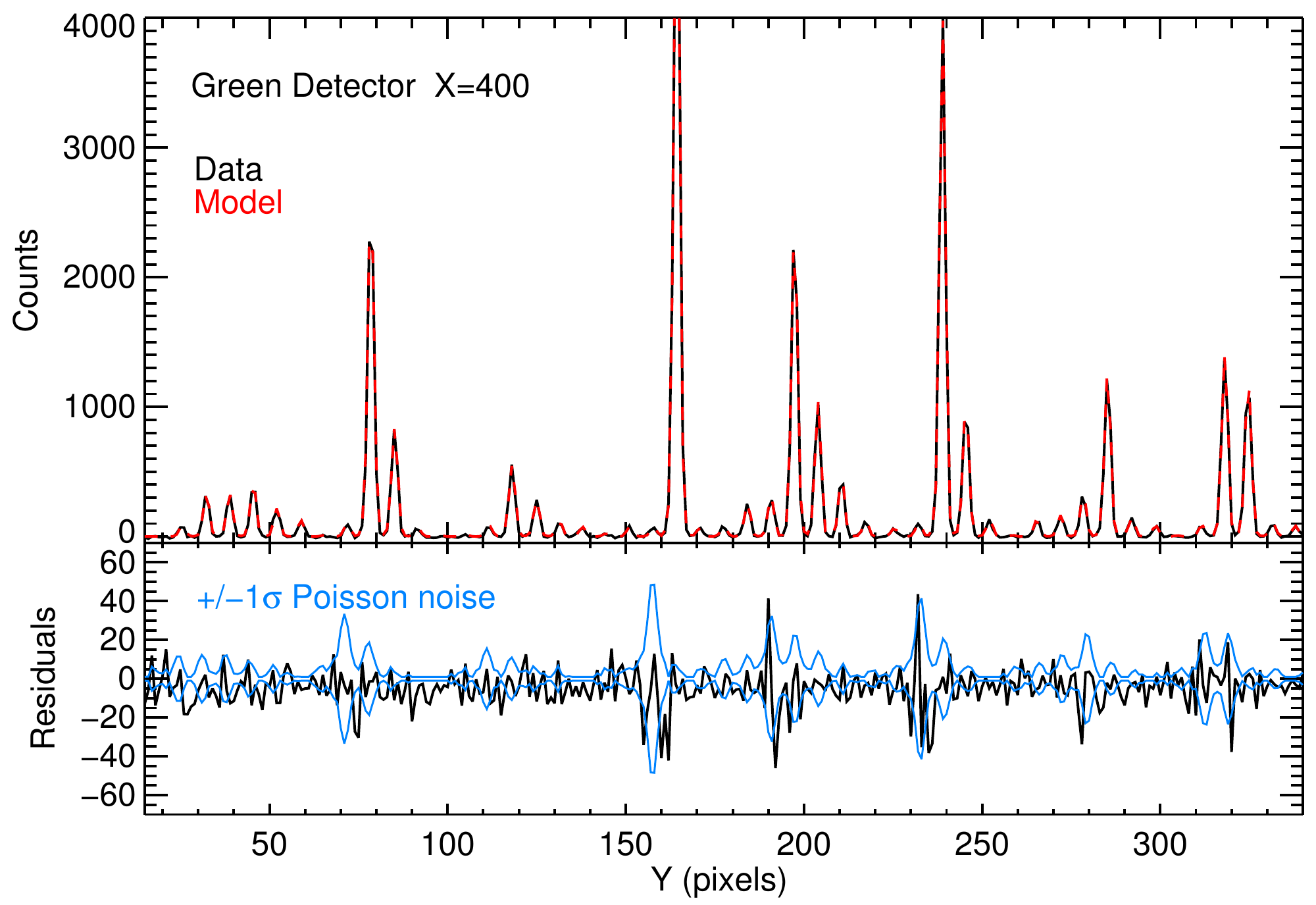}
\end{center}
\caption{An example of the PSF model (red) of a stellar spectrum (black) along a section of one column (X=400)
  in the green detector.  The residuals (data$-$model) are shown in the lower panel as well as $\pm$
  the Poisson noise indicating that the model is satisfactory.}
\label{fig_psfmodel_example}
\end{figure}

\begin{figure}[b!]
\begin{center}
\includegraphics[angle=0,scale=0.28]{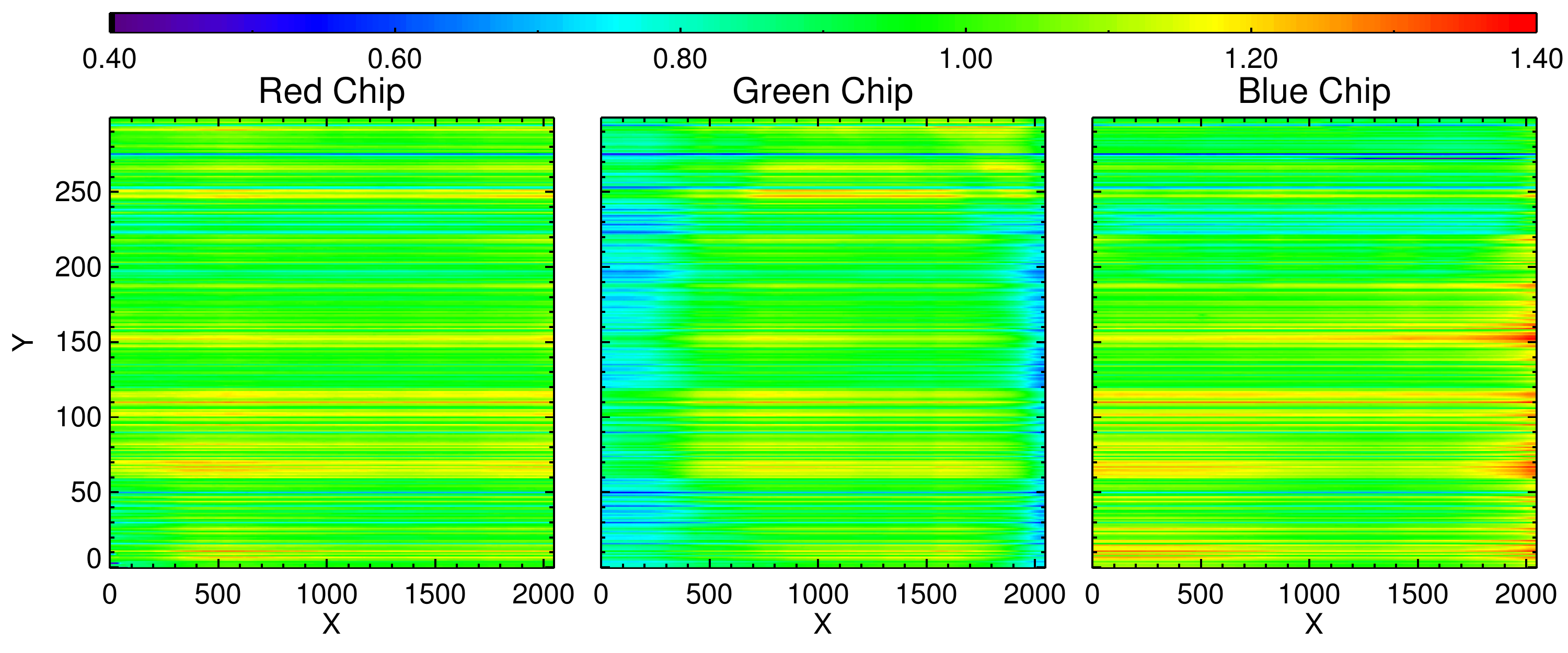}
\end{center}
\caption{An example of fiber-to-fiber throughput variation calibration functions for all 300 fibers
  and three detectors.  The rms variations are 9.7, 12.2, and 11.3\% for the red, green and blue
  detectors respectively.}
\label{fig_fluxcorr}
\end{figure}

From the ``dome'' flats, a rough correction for scattered light is
made by subtracting the mean value measured at the top and bottom
of the chips (specifically, rows 5--10 and 2038--2042). The resulting image is then smoothed
in the wavelength direction with a boxcar filter of width 50 pixels,
to increase $S/N$, given that the PSF is not expected to change significantly on
this scale (there is very little tilt of the traces against detector rows).
At each column, the observed profile is then tabulated
for each of the 300 traces. Since there is some overlap between
adjacent traces, the light at each pixel is distributed between
the two surrounding traces using information from 
the sparse pack calibration frames.
Based on the distances of a given pixel from each of the surrounding traces,
the sparse pack calibration is used to determine the relative fraction
of the two contributions by looking at values at these distances
(on the appropriate side of the PSF)
in the nearest trace in the tabulated sparse pack data.

This method assumes that the profile is sufficiently limited in 
spatial extent so that only adjacent traces contribute to the light
at any given pixel. With this assumption, the brightness of any
given trace is directly coupled only to the brightness of adjacent
traces.  Since the solutions of neighboring triplets of fibers are 
linked together, the solution of all fiber fluxes becomes
a linear algebra problem with a tridiagonal matrix, solving for
300 individual fluxes at each column.  This problem 
is solved by using the tridiagonal matrix or ``Thomas'' algorithm \citep{Thomas49}.
Errors are propagated through the extraction.
An example of a model PSF fit to the APOGEE data is shown in Figure \ref{fig_psfmodel_example}.

\begin{figure}[t!]
\begin{center}
\includegraphics[angle=0,scale=0.53]{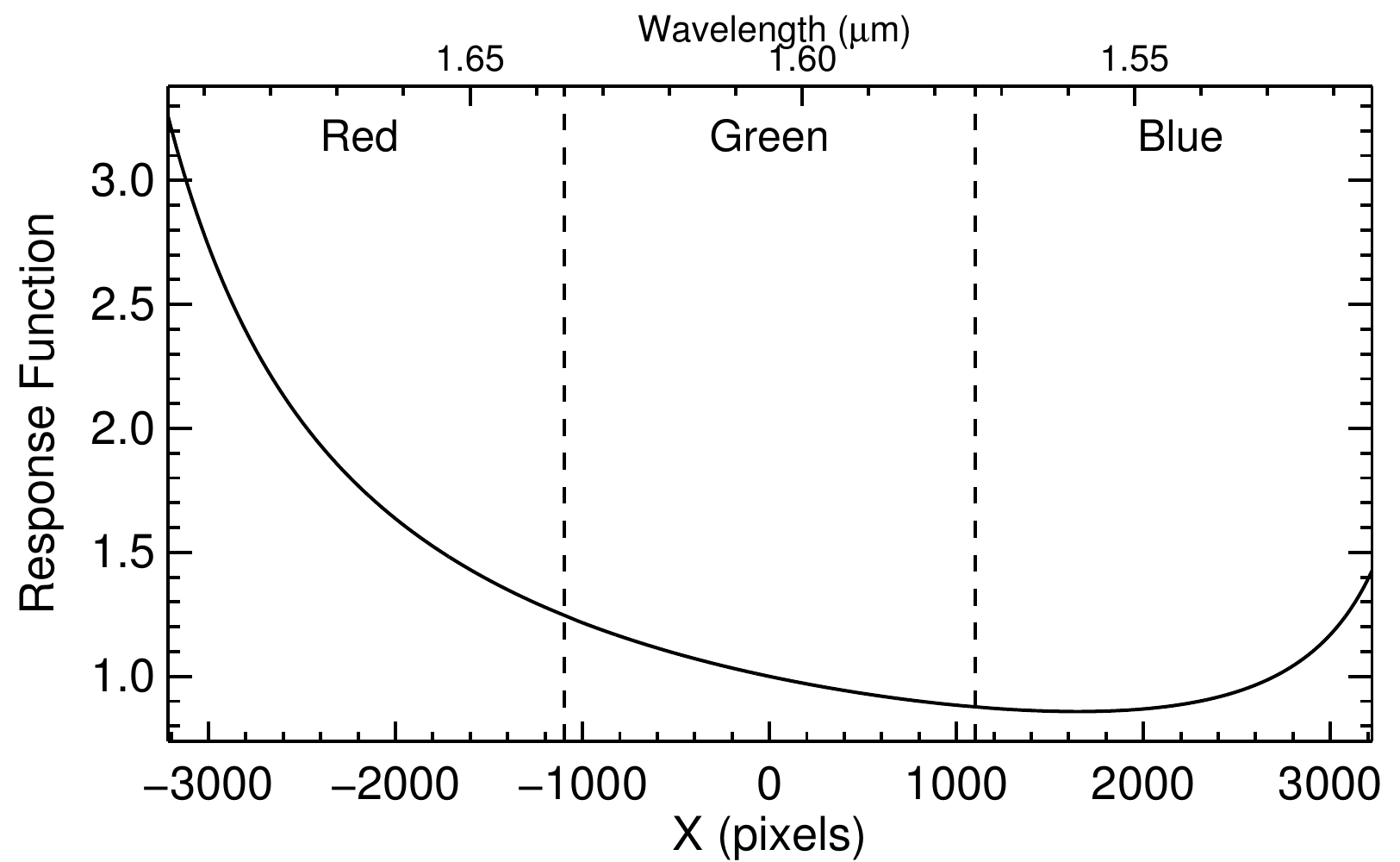}
\end{center}
\caption{The response curve calibration spectrum.  Each spectrum is multiplied by
this function to roughly calibrate the relative fluxes.}
\label{fig_responsecorr}
\end{figure}



The AP2D code also includes options for three other
extraction methods: boxcar extraction, and simultaneous fitting
using both a simple Gaussian and a more complex functional (but still
parametric) form for the PSF. Future work might include developing
these further. The most significant future development, however, would
probably be to go to a full two-dimensional extraction that accommodates
a 2D PSF that is not separable in rows and columns \citep[i.e., ``spectro-perfectionism'',][]{Bolton10}.




\subsection{Fiber-to-fiber throughput variation and response curve correction}
\label{subsec:fluxcalib}

The ``dome'' flats for each plate are then used to correct
the 300 extracted spectra for moderately wavelength-dependent fiber-to-fiber throughput
variations. These are measured on an individual plate basis because
of the possibility that the relative throughputs vary depending
on the details of each specific coupling of the long fibers
from the instrument to the short fibers in the plug plates via the gang connector.
Figure \ref{fig_fluxcorr} shows an example of a flux calibration file.  The fiber-to-fiber
variations are at the $\sim$10\% level.


Accurate fiber-to-fiber throughput correction is desired to
enable the possibility of accurate sky subtraction using
separate sky fibers.  The correction reduces the rms variation
of the flux in sky fiber lines across the plate from $\sim$12\% to $\sim$5\% and
only $\sim$1\% around a smoothly varying 2D spatial polynomial fit to the flux
variations.


After the fiber-to-fiber throughput calibration, each spectrum
is corrected by a wavelength-dependent spectral response function
to apply an approximate relative flux calibration.  The relative response
curve was created using the spectrum of a blackbody of well-known
temperature (110 $^{\circ}$C and 150 $^{\circ}$C) and is used for all APOGEE spectra (Figure \ref{fig_responsecorr}).

%

\subsection{Wavelength Calibration}
\label{subsec:1dwavecal}



The correspondence between wavelength and pixel is determined using exposures of
ThArNe and UNe hollow-cathode lamps. The wavelength solution varies from fiber-to-fiber because
of the optical design and also because of the exact placement of each fiber in
the pseudo-slit. The former causes variations over large scales, while the latter
causes a fiber-to-fiber shift.

\begin{figure}[t]
\begin{center}
\includegraphics[angle=0,scale=0.38]{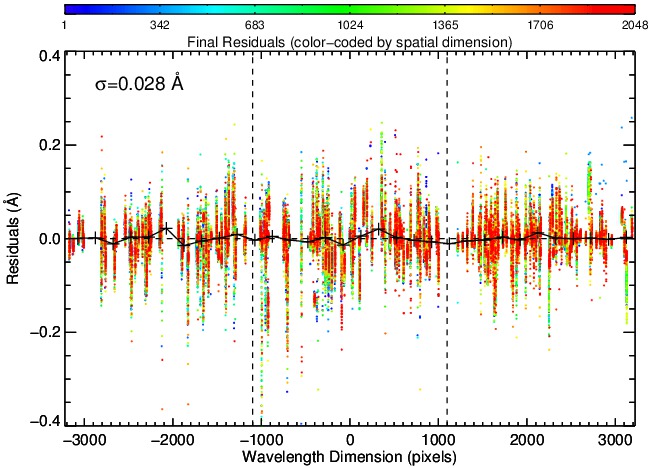}
\end{center}
\caption{Residuals of Thorium-Argon-Neon and Uranium-Neon line fits around the wavelength solution
  versus position along the three detectors (color-coded by position in the spatial dimension).
  The global rms is 0.028\AA~and dominated by the measurement uncertainty
  in the line centers.  Connected black crosses show medians in bins of 200 pixels.  Systematics in
  the binned residuals are at the $\sim$0.01--0.02\AA~level.}
\label{fig_wavelength_resid}
\end{figure}

To derive the wavelength solution, Gaussians are fit to all detectable lines (4$\sigma$ above
the background) in all calibration spectra.  Next, these Gaussians are matched up to known lines.  The information for
all the lines in the 300 fibers, three arrays, and multiple exposures (normally one ThArNe and one UrNe)
are combined into one list.  A 5th order polynomial ($\lambda$ versus pixel) and two detector gaps are fit
to each fiber separately.  A robust linear fit (to allow for the possibility of small rotations of the
detectors relative to each other) is made to the derived detector gaps as a function of fiber number to obtain
more accurate values.  Finally, the 5th order fits are redone holding the chip gaps
fixed at the fitted values.  The residuals to these fits are on the order of
$\sim$0.03--0.04\AA~($\sim$0.1 of a pixel) and dominated by the Gaussian line-fitting errors.
Systematics in the residuals are at the $\sim$0.01--0.02\AA~level (Figure \ref{fig_wavelength_resid}).
The temporal variations over year timescales is $\sim$0.015\AA, which demonstrates that the instrument
is very stable.

A wavelength solution from an apWave calibration file is applied
to each extracted spectrum.  This wavelength solution still needs to
be corrected for slight differences between the science exposures and
the calibration frames due to different dither positions and potential changes
in the optics over time (see \S \ref{subsec:lsfcal}).  For ``on-sky''
observations (normal science exposures, sky flats, and Any Star
Down Any Fiber [ASDAF]\footnote{ASDAF exposures are special exposures of bright stars (often ``standard'' or calibration stars)
placed on a specific fiber.} exposures) the night sky airglow emission
lines are used for this correction.  A zero-point offset in the
pixel positions (not the wavelength) is determined for each fiber
separately.  First, Gaussians are fit to all the bright emission
lines in each spectrum.  Then, a first guess zero-point offset is
determined by cross-correlating a model airglow spectrum (using
Gaussians with heights of unity and a standard APOGEE wavelength
solution) with a model spectrum of the measured lines (also using
Gaussians with heights of unity).  Next, the zero-point estimate
is used to match the measured lines with known airglow lines.  A
new wavelength solution is determined allowing only the pixel
zero-point to vary.  Finally, a line is fit to all the fiber
zero-point shifts versus fiber number and this fit is used for the
zero-point offsets in the wavelength solution.


\subsection{Output}

The output files for AP2D are called ap1D-[abc]-ID8.fits and have four extensions containing the
flux, errors, bitwise pixel mask, and wavelength array (in \AA), each  with a size of 2048$\times$300.
A model of the 2D image is also output with names of ap2Dmodel-[abc]-ID8.fits.  These files 
are output to the {\tt spectro/v\#/red/MJD5/} directory.

\section{AP1DVISIT: Visit Stage}
\label{sec:ap1dvisit}


After extraction, the spectra are sky and telluric corrected and then the separate
exposures (at different dither positions) are combined into one well-sampled spectrum per fiber.
This is performed in the AP1DVISIT stage with the AP1DVISIT routine.

As explained in Section \ref{sec:operations}, each plate ``visit'' generally consists of eight $\sim$500 sec (47-read) exposures taken
as two ABBA dither position sequences (A and B being roughly 0.5 pixels apart in the spectral
dimension).

Sky subtraction and telluric absorption correction are done on an exposure-by-exposure
basis because of the possibility of sky variation from one exposure to another, since
the sky is known to vary on short time scales. This leads to some complications because
of the mild undersampling of the spectra on the blue end, preventing simple resampling
of sky and telluric fibers to yield a correction for each object fiber.



\subsection{Dither Shift Measurement}
\label{subsec:dithershift}

Due to slight undersampling at the bluer wavelengths the APOGEE exposures are taken at two
different spectral dither positions
shifted by roughly 0.5 pixels from each other.  The actual positions at which the exposures
are taken are not known precisely enough a priori for accurate combination of spectra taken at two
different dither positions.  Therefore, the dither position of each exposure is measured from
the actual data.  This is done relative to the first (reference) exposure
in two ways:
(1) cross-correlation
of spectra, and (2) shifts of airglow emission lines.  For the cross-correlation, the
(non-sky) spectra are first normalized (using a 100-pixel median filter), then cross-correlated
against the first exposure spectrum, and finally a Gaussian is fit to the cross-correlation peak
to find the best shift.  This is done separately fiber-by-fiber and array-by-array.  Finally,
a robust mean is calculated of all individual 900 shifts measured.  In the emission line method,
all bright emission lines in all fibers are fit with Gaussians, then matched with their corresponding
lines in the first exposure, shifts calculated, and a robust mean measured.  The pipeline currently
uses only the cross-correlation technique which gives average formal errors of $\sim$0.005 pixels.
The measured dither shifts are written to the header for later use in the dither combination process.

\begin{figure}[t!]
\begin{center}
\includegraphics[angle=0,scale=0.45]{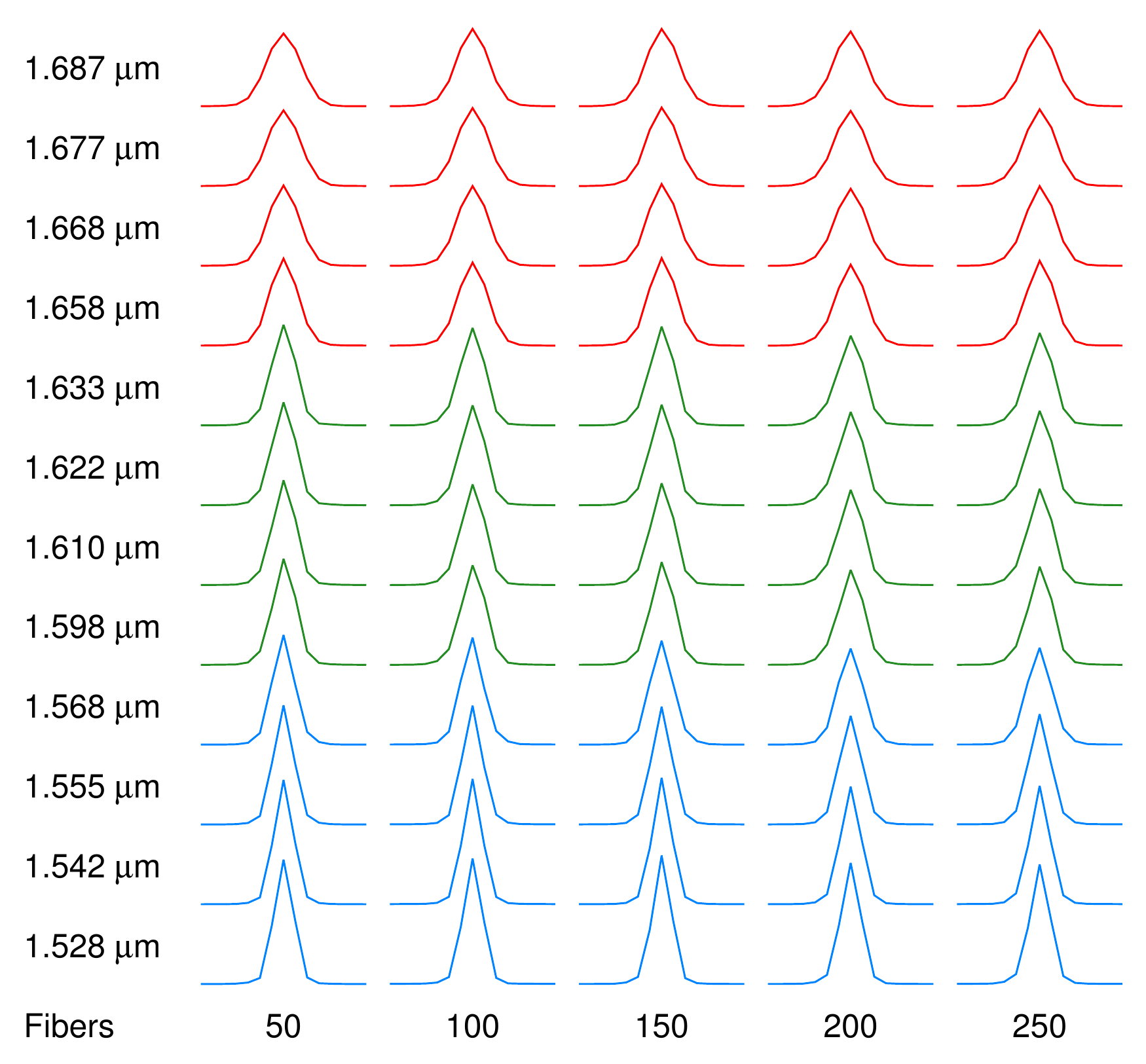}
\end{center}
\caption{A diagram illustrating the LSF variations across the detectors.  Each profile
  shows $\pm$7 pixels of the model LSF convolved with the pixel size.}
\label{fig_lsfdiagram}
\end{figure}


\subsection{Line Spread Profile (LSF)}
\label{subsec:lsfcal}

To accomplish sky correction (emission and absorption) on undersampled data, as
well as for spectra in which the LSF varies across the detector, the sky
corrections are forward modeled using constraints from the observed sky
and telluric fibers. This requires an accurate measurement of the LSF as
a function of wavelength and fiber.

The line spread function (LSF) is modeled as a sum of Gauss-Hermite functions\footnote{We use the
probabilists' polynomials, which are given by: $He_n(x)=(-1)^n e^{\frac{x^2}{2}} \frac{d^n}{dx^n} e^{-\frac{x^2}{2}}$.
\path{https://en.wikipedia.org/wiki/Hermite_polynomials}} (which form an orthonormal basis) and
a wide-Gaussian for the wings.  The convolution of the model LSF with the pixel
size is taken into account. LSF parameters are determined using airglow lines in the nightly skyflats.
Each fiber and array are fit separately.
The LSF parameters are allowed to vary slowly with wavelength by means of a low-order
polynomial as a function of pixel.  The model LSF is always normalized to an area of one.
Figure \ref{fig_lsfdiagram} illustrates the typical variations of the LSF across the three
detectors and the fact that the LSF is slightly undersampled in the blue.

\begin{figure*}[t!]
\begin{center}
\includegraphics[angle=0,scale=0.55]{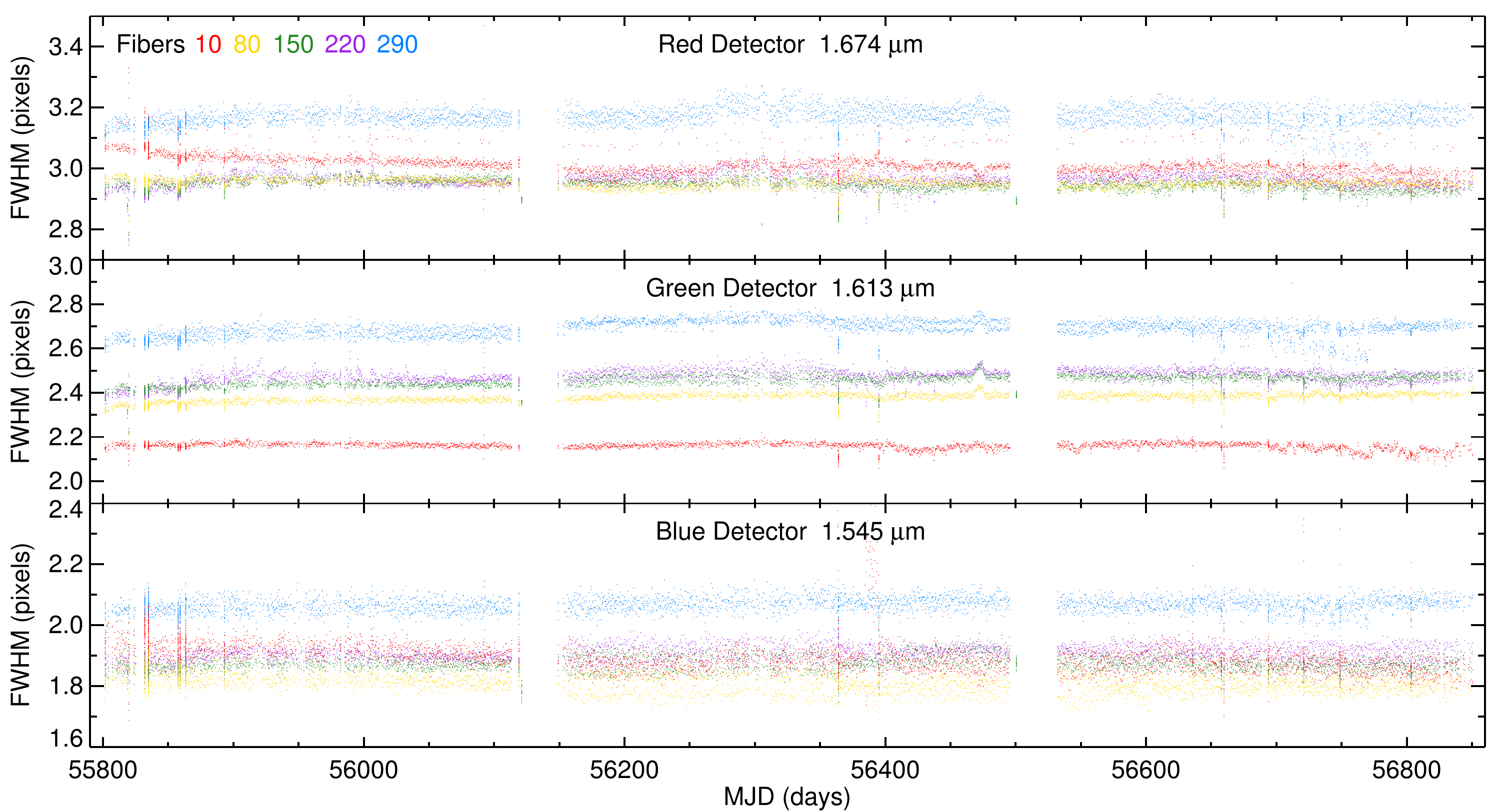}
\end{center}
\caption{A demonstration of the temporal stability of the LSF.  The Gaussian FWHM (in pixels) of one ThArNe line per detector
  and five fibers versus MJD.  The FWHM values are stable at the $\sim$1--2\% level over three years.}
\label{fig_lsfstability}
\end{figure*}

As a demonstration of the LSF stability, Figure \ref{fig_lsfstability} shows the temporal
variation of the Gaussian FWHM of one Thorium-Argon-Neon line per detector (and five fibers).
While there are some small-scale variations, the rms of the FWHM values per fiber are
$\sim$1--2\%, which indicates that the instrument has been remarkably stable over three years.

\begin{figure*}[t!]
\begin{center}
\includegraphics[angle=0,scale=0.59]{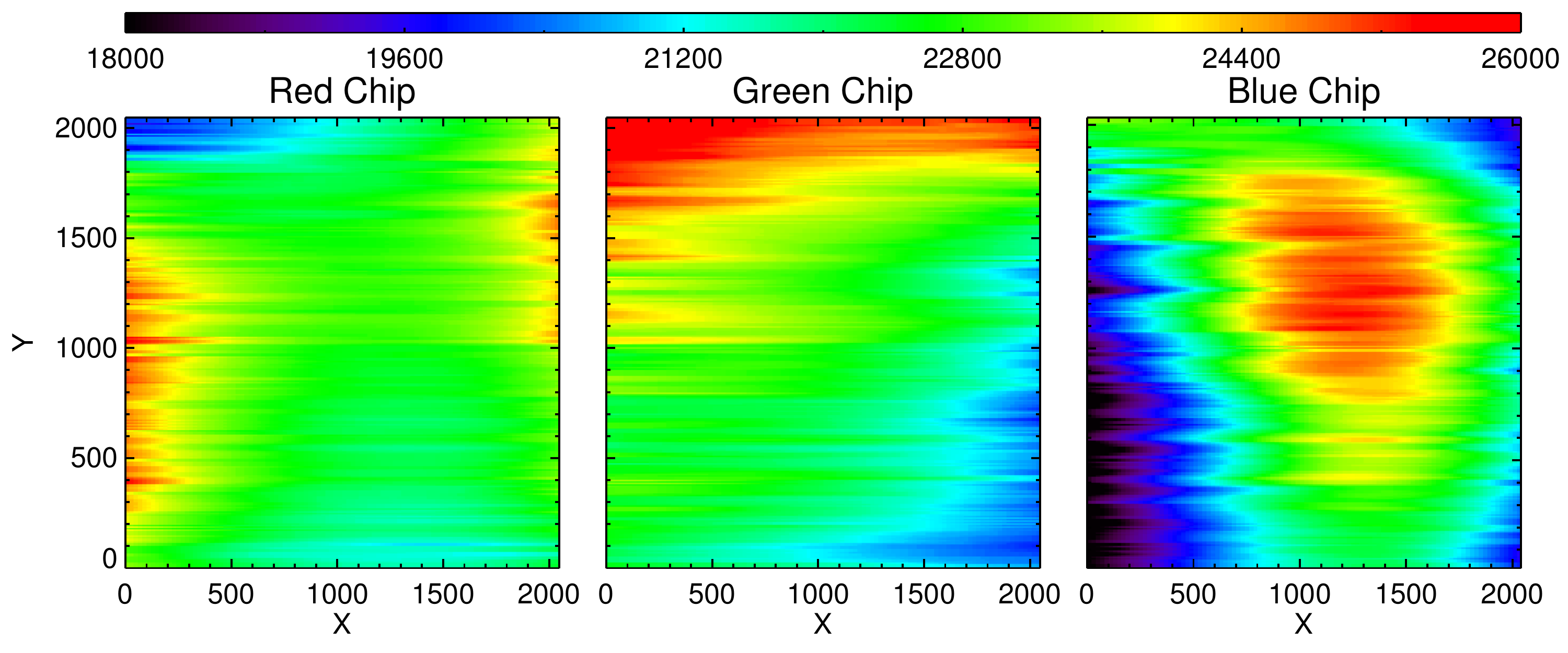}
\end{center}
\caption{Maps showing how the resolving power (R=$\lambda$/$\Delta$$\lambda$) varies across the three detectors.}
\label{fig_resolution}
\end{figure*}

The traditional measurement of resolving power (R=$\lambda$/$\Delta$$\lambda$; using a direct-measured FWHM for $\Delta$$\lambda$)
gives an average of R=22,500 for APOGEE spectra, but there are $\sim$10--20\% variations across the detectors (spectrally
and spatially) as seen in Figure \ref{fig_resolution}.  However, the APOGEE LSF is non-Gaussian and, therefore, a
traditional ``R''-value is not necessarily a good description of APOGEE's resolution.







\subsection{Sky Subtraction}
\label{subsec:skysub}

The observed spectra are contaminated by night sky emission airglow lines (mostly OH) from the Earth's
atmosphere, and sky continuum which is most often dominated by moonlight (reflected sunlight)
and greatly enhanced on cloudy nights.  We have found that light pollution from nearby
El Paso (Texas) is not a significant component of our sky spectra.

The sky spectrum can vary spatially across our 3\dgr FOV and temporally during our $\sim$1 hour ``visit''
of the field.  Therefore, each plate has $\sim$35 fibers designated for ``blank'' sky that can be used to
subtract the sky spectrum from the science fibers.  The sky fiber positions are chosen in $\sim$17 spatial
zones to give them a fairly uniform distribution across the plate;
see \citet{Zasowski2013} for details.

\begin{figure*}[ht!]
\begin{center}
\includegraphics[angle=0,scale=0.65]{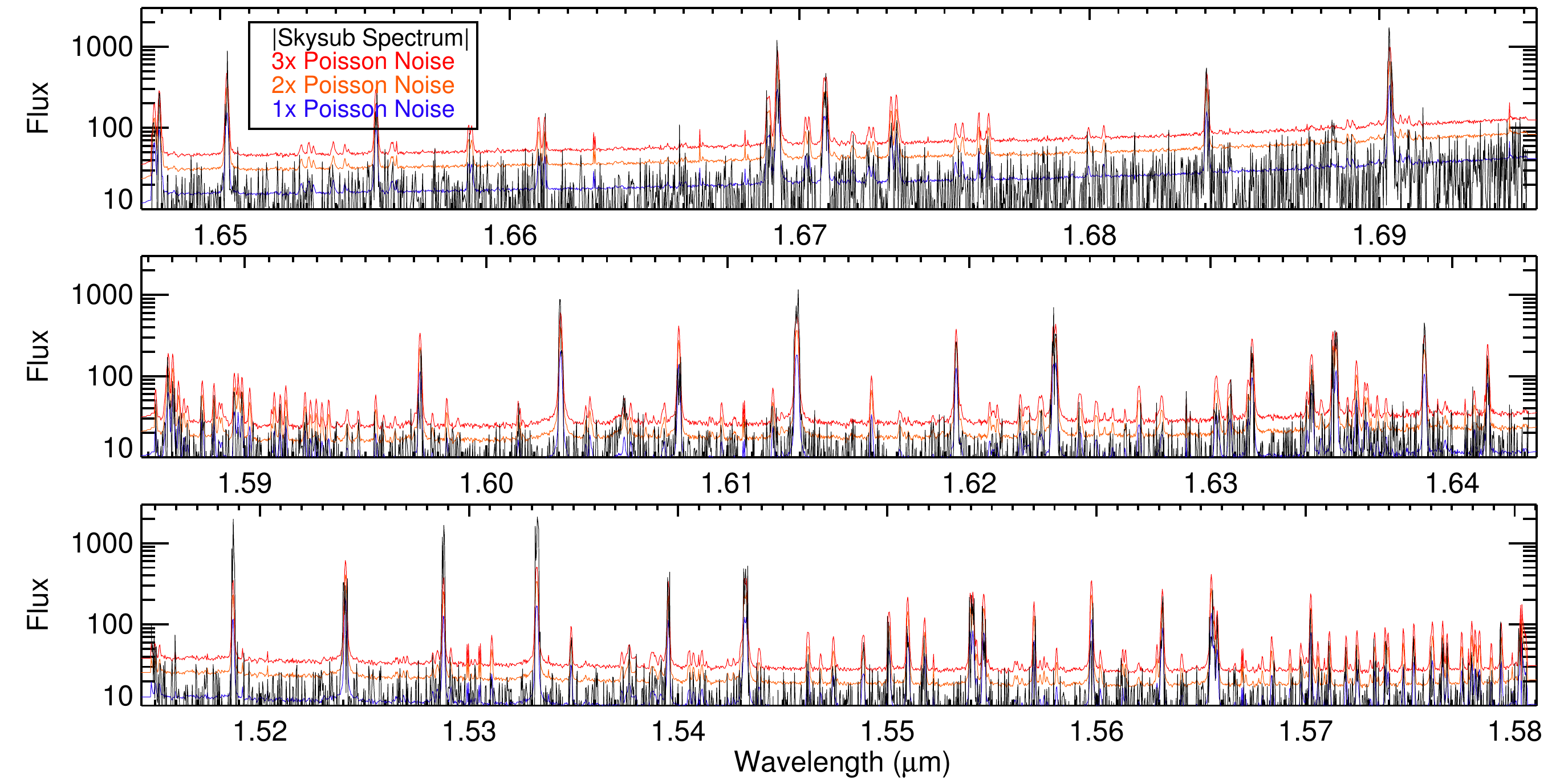}
\end{center}
\caption{Example of a single exposure APOGEE sky fiber spectrum (exposure 02870014, fiber 258) after sky subtraction
(the absolute value; the sky fiber in question was not used to determine the sky spectrum, only the
neighboring sky fibers were) with the pipeline noise model for comparison.  Most sky line residuals
are below 3$\times$ the noise model.}
\label{fig_skysub}
\end{figure*}

The current pipeline is using a temporary, and sub-optimal, sky subtraction method.
For each science fiber the four nearest (on the sky) sky fibers are found.  First, the spectra from
these four sky fibers are resampled onto the wavelength solution of the science fiber (using cubic
spline interpolation).  Next, each of the four sky spectra are cross-correlated with a continuum
subtracted (150 pixel median filter) science fiber spectrum and shifted accordingly to fix any
errors in the wavelength solutions.  An emission line scaling factor is then calculated for each
sky spectrum relative to the science spectrum airglow lines using pixels with sky emission greater
than 10$\times$ the noise.  A weighted average sky spectrum is then created from the scaled sky
spectra (with sky continuum added back in) using 1/(sky distance)$^2$ for the weighting, and
subtracted from the science spectrum.  Figure \ref{fig_skysub} shows an example of a single exposure sky fiber
spectrum after sky subtraction (the absolute value; the sky fiber in question was not used to
determine the sky spectrum, only the neighboring sky fibers were) and the pipeline noise model for comparison.
Even this sub-optimal sky subtraction works fairly well with the large majority of the residuals
below three times the noise level.

Possible improved methods include modeling of the airglow lines with 2D spatial polynomial fitting
of airglow family fluxes, and principal component analysis (PCA). These methods are currently under
investigation and development.

It should be noted that the airglow lines are so bright that little scientific benefit can be gained
from the stellar spectrum ``underneath'' them, since even with perfect subtraction the Poisson
noise from airglow will dominate over (wash out) any stellar spectrum.  The far wings of the airglow lines should be
salvageable, but the most critical portion of sky subtraction is the removal of the sky continuum, which would
otherwise distort line depths in normalized spectra and their subsequent analysis.  Poorly subtracted
airglow lines can also adversely affect the RV determination.

\subsection{Telluric Correction}
\label{subsec:telluric}

\begin{figure*}[ht!]
\begin{center}
\includegraphics[angle=0,scale=0.35]{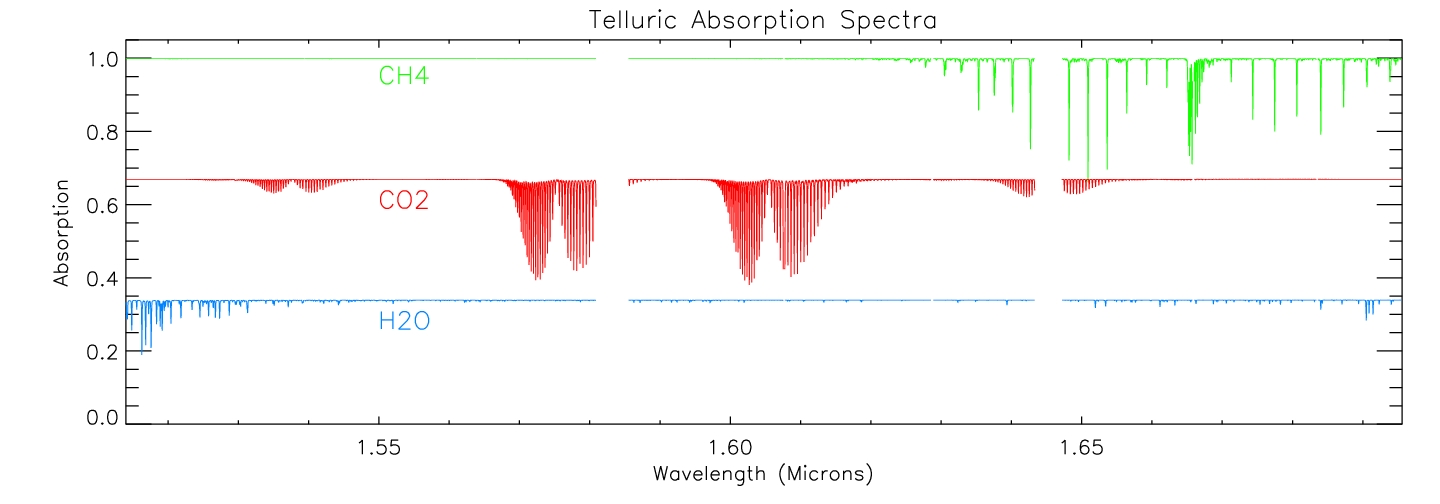}
\end{center}
\caption{The atmospheric telluric absorption features in the APOGEE spectral window due to
CH$_4$ (green), CO$_2$ (red) and H$_2$O (blue).}
\label{fig_telspec}
\end{figure*}

The NIR $H$-band hosts a number of atmospheric (telluric) absorption lines (from 
H$_2$O, CO$_2$, and CH$_4$; see Figure \ref{fig_telspec}) that contaminate a significant
fraction of the APOGEE spectral range ($\sim$20\%) and need to be corrected.  As with the airglow spectrum
the telluric absorption spectrum may vary spatially and temporally for our observations.
Each plate has $\sim$35 fibers designated for ``telluric'' (hot) stars that can be used to ascertain
the telluric absorption for the science fibers.  A procedure is used to pick telluric fibers in spatial
zones similar to that used for sky fibers \citep{Zasowski2013}.

A three-step process is used to correct for telluric absorption:
(1) telluric absorption model fitting to the hot star spectra, (2) 2D polynomial spatial fitting
of the telluric species scaling parameters across the plate, and (3) construction of the model telluric
absorption spectrum for each science fiber using the model telluric spectra, 2D polynomial fitting parameters,
and the known LSF of the science fiber.

The LBLRTM\footnote{\path{http://rtweb.aer.com/lblrtm.html}} model atmosphere code \citep{Clough05} was used to create
a grid of high-resolution model telluric species spectra for CO$_2$, H$_2$O and CH$_4$, individually, using the
US Standard 1976 atmosphere for the altitude of APO. For
CO$_2$ and CH$_4$, four different scale factors that parameterize the strength of the features were used 
(0.5, 1.0, 1.5, and 2.0 cm); for H$_2$O, four different precipitable water columns (0.75, 1.5, 2.25, and 3.0) 
were used. Spectra were calculated at seven different airmasses, from 1.0 to 2.5, spaced by 0.25.

For each plate, we interpolated in airmass within this grid to get four model spectra for each absorption
species. For each telluric star on a plate, we find the scale factor of each model that best matches
each spectrum, and determine which of the four scaled models for each species provides the best fit. We then
adopt the model for each species that best fits the majority of the stars,
and refit each telluric spectrum with the same model to get a self-consistent set of scale factors
across the field. 

The fits are performed by adopting the scale factor that yields the minimum RMS in all pixels
where the telluric lines for that species are dominant (pixels must
be within 5 pixels of a telluric species line with strength greater than 1\% in the
convolved model telluric spectrum).  To obtain a reliable telluric scale factor the stellar continuum
(including the wide hydrogen absorption lines) needs to be removed.  Therefore, an iterative process is used.
The stellar continuum is calculated using a median filter (100 pixels wide) of the stellar spectrum
corrected with the current best-fit model telluric spectrum (after the first iteration).  Stellar continuum
and scale factors are calculated until the solution converges.
Since the line spread profile (LSF) varies from fiber to fiber,
the high-resolution model species spectra are convolved with a separate LSF (from the apLSF
calibration file) for each fiber.

\begin{figure}[t!]
\begin{center}
\includegraphics[angle=0,scale=0.35]{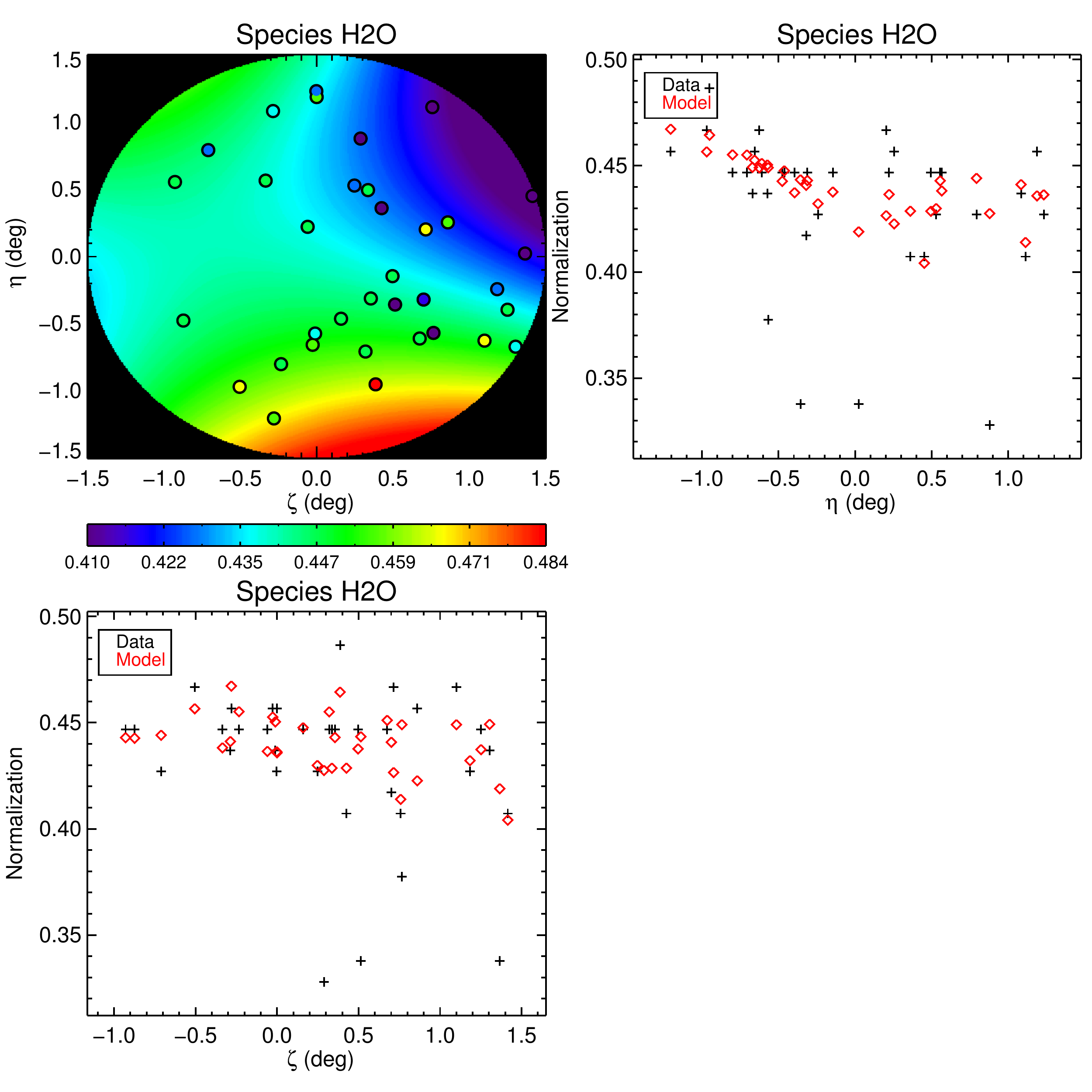}
\end{center}
\caption{Example of a 2D polynomial fit to the telluric species scaling for H$_2$O (exposure 03770010).  $\zeta$ and $\eta$ are
coordinates relative to the center of the plate in right ascension and declination, respectively.
This is a fairly extreme case to show how the 2D fitting handles large spatial variations.  Most cases are flatter.}
\label{fig_telsky}
\end{figure}

At this point the species scalings for each hot star are determined, and a 2D spatial polynomial model
is fit to the $\sim$35 scalings for each species separately. For CO$_2$ and CH$_4$ we use
a linear model; since these species are thought to be well-mixed in
the atmosphere, the linear trend is included to account for variations in airmass
across the field. For H$_2$O, which could have significant spatial structure, we use
a quadratic surface.
The RMS around the fit is normally $\sim$1--2\%.  Figure \ref{fig_telsky} shows an example of a 2D
polynomial fit to a species scaling with significant spatial variations (most exposures show much less variations).


Finally, the model telluric spectra are calculated for each science spectrum, using the 2D polynomial 
fit coefficients to obtain the three species scalings for the position of the science object.  Then, the high-resolution
model telluric species spectra are convolved with the LSF of the science fiber and scaled appropriately with
the species scaling for that fiber.
The science spectrum is divided by the final convolved telluric spectrum.
An error in the telluric model for each object is computed by taking the RMS scatter in the 2D polynomial fit of
each species scaling and propagating the errors forward into the model telluric spectrum. The average error in the
telluric correction is roughly $\sim$1--2 \% of the stellar continuum.  It is worth noting that this procedure is
astrophysically incorrect as the telluric absorption occurs {\em before} the light goes through the spectrograph and
gets convolved with the instrumental LSF.  However, this often-used approximation works well enough ($\sim$1\%) for
our needs.

\begin{figure*}[ht!]
\begin{center}
\includegraphics[angle=0,scale=0.59]{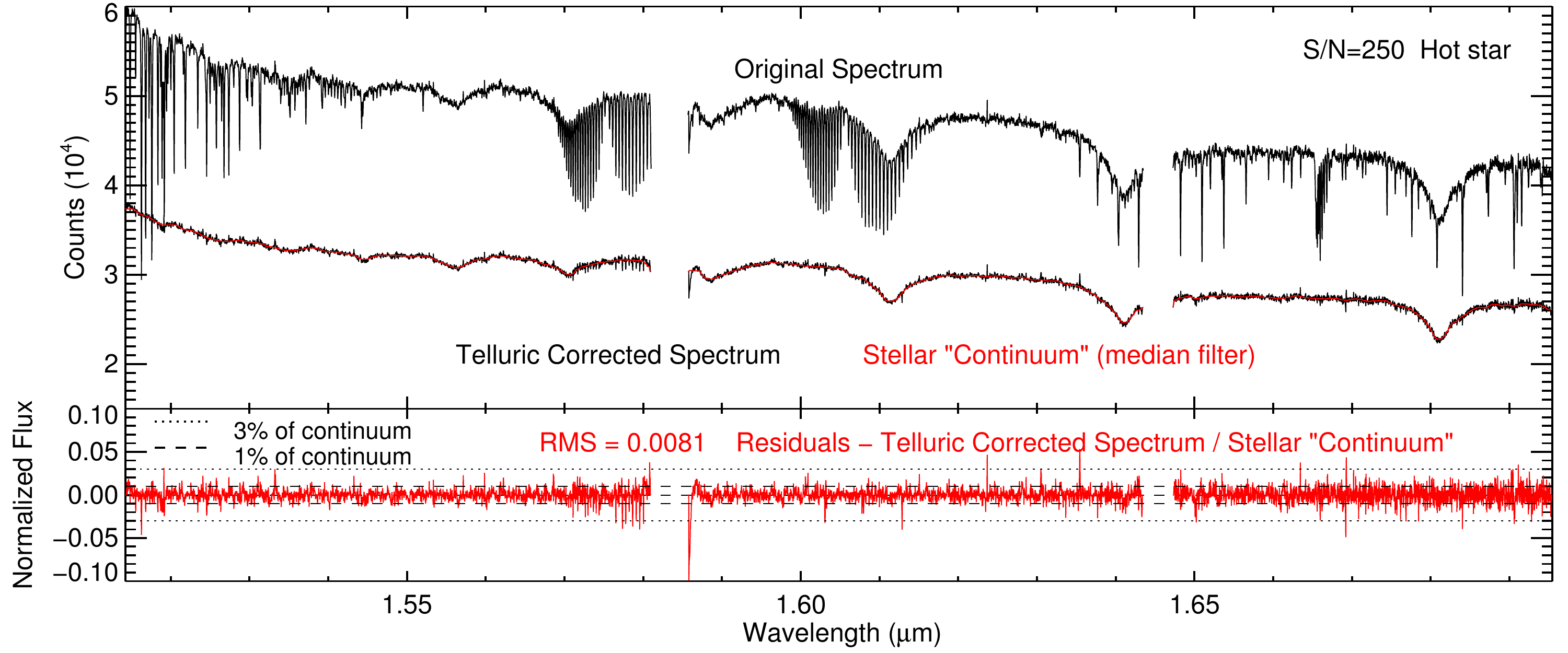}
\end{center}
\caption{Example of an APOGEE spectrum of a hot star before and after telluric correction.  (Top) Original spectrum
in counts with telluric corrected spectrum below and the median-filtered stellar ``continuum'' (red). (Bottom)
The ``residuals'' (red; telluric corrected spectrum divided by the stellar ``continuum'' minus 1) with an RMS of 0.0081.}
\label{fig_telcorr}
\end{figure*}

Figure \ref{fig_telcorr} shows an example of an APOGEE spectrum of a hot star before and after telluric correction.  The
residuals after removing the broad stellar absorption features are small at 0.0081 or less than a percent.  Some systematics
in the residuals, such as around the CO$_2$ band on the red side of the blue chip, are still present.
Since the LSF varies only slowly with time it is not necessary to convolve the high-resolution telluric model spectrum
for each exposure separately.  Therefore, to save time in the entire telluric correction step, the three original
high-resolution telluric model spectra are pre-convolved with the LSF of all 300 fibers and saved as the apTelluric
calibration product.
Currently, only a single LSF calibration file is used for all the APOGEE reductions, because the LSF shows very little temporal
variation (see Figure \ref{fig_lsfstability}).

After sky subtraction and telluric correction the ``corrected'' frame is written to disk.  The apCframe-[abc]-ID8.fits
files have extension of 1-flux (electrons), 2-error (electrons), 3-bitwise flag mask, 4-wavelength (\AA),
5-sky (electrons), 6-sky error (electrons), 7-telluric absorption, 8-telluric error, 9-wavelength
coeficients, 10-LSF coefficients, and 11-plugmap structure (binary table).  The arrays in extensions 1--8
have sizes of 2048$\times$300, the wavelength coefficients 300$\times$14, and LSF coefficients 300$\times$27.

\subsection{Dither Combination}
\label{subsec:dithercomb}

As mentioned in \S \ref{subsec:dithershift}, due to undersampling the APOGEE exposures are taken
at two different dither positions that must be combined to create well-sampled spectra.  This is
performed in two steps: (1) the exposures (a total of N$_{\rm exp}$) are paired up and interlaced to create well-sampled
spectra, and (2) the N$_{\rm exp}$/2 well-sampled spectra are co-added to create one ``visit'' spectrum
per object.

The exposures are paired up, one dither position (A and B) per pair to create a series of approximately equal-$S/N$
pairs.  Each exposure spectrum is separately normalized/scaled using a median filtered
(width of 501 pixels) version of the spectrum
(each array separately) to remove any variations in net flux due to differences in seeing or throughput between the two
exposures (only for objects, not sky).  Using the previously measured dither shifts (\S\ref{subsec:dithershift}), 
the two scaled spectra are then combined with the sinc-interlace equations from \citet{Bracewell}
onto a pixel scale twice as fine as the ``native'' scale (for the single exposures).  All pairs share the same
final pixel scale so that the spectra are only resampled once.
The flux level of the well-sampled spectrum is then rescaled to the average of the scaling arrays of the two
individual exposures.

In the second step, the N$_{\rm exp}$/2 well-sampled spectra (normally 4 per visit) are co-added.
The flux level of the spectra are again scaled using a median filter.  The scaled spectra are combined using
a weighted mean (with 5 $\sigma$ outlier rejection) where the weights are computed either on
a spectrum-to-spectrum or pixel-by-pixel basis.  The pipeline currently uses the pixel-by-pixel
weighting.  The final combined spectrum is then rescaled using the sum of the scaling arrays of the individual spectra.




\subsection{Absolute Flux Calibration}
\label{subsec:absfluxcal}

The final combined object spectra are roughly flux calibrated (on an absolute flux scale)
using the 2MASS $H$-band magnitudes.  For each star we separately determine an absolute
flux correction factor, $c_{abs}$, to allow for differences in airmass, fiber-centering errors,
and seeing variations.  This multiplicative factor is used to convert the observed spectrum
in electrons to calibrated physical units,


\begin{equation}
F_{\lambda} = c_{abs} S_{i}
\end{equation}

\noindent
where $S$ is the observed spectral flux in e$^{-}$ at pixel $i$.  The correction factor can be calculated with,

\begin{equation}
c_{abs} =  F_{\lambda,0} 10^{-H/2.5} / MEDIAN(  S_{i} )
\end{equation}

\noindent
where $F_{\lambda,0}$ is the 2MASS zero-point for isophotal monochromatic light (1.33$\times$10$^{-13}$ W/cm$^2/\micron$)
for a 0th magnitude star in the $H$-band
\citep{Cohen03}\footnote{\path{http://www.ipac.caltech.edu/2mass/releases/allsky/doc/sec6_4a.html} }.
After applying the absolute flux calibration the spectra are in physical units of ergs/s/cm$^2$/\AA.

\subsection{Output}

After the absolute flux calibration step
the apPlate-[abc]-PLATE4-MJD5.fits and apVisit-PLATE4-MJD5-FIBERID3.fits
files are written to disk.  The apPlate files contain all 300 spectra while the apVisit files ($\sim$265 of them)
are for single object spectra (no sky spectra).  The fluxes and errors are stored in units of
10$^{-17}$ ergs/s/cm$^2$/\AA.


Note, the FiberID is not the same as the IDL index in the data arrays.  FiberID=300 is the first spectrum in
the data arrays (bottom) and FiberID=1 is the last one (top).  The conversion is FiberIndex=300$-$FiberID.

Many of the important object parameters including name, coordinates, 2MASS magnitudes,
APOGEE targeting flags, date observed, apVisit filename, radial velocity, best-fitting
template parameters, and more are saved in a FITS binary table called apVisitSum-PLATE4-MJD5.fits.
in the field directory {\tt spectro/v\#/fields/LOC4/}.  These are useful for quickly
accessing the output parameters of the pipeline.

\section{APSTAR: Object Stage}
\label{sec:apstar}


Most of the stars are observed in several different visits to enable
detection of radial velocity variation to identify binaries and to
accumulate the necessary $S/N$. After multiple visits, a combined spectrum
is made to provide the highest $S/N$ individual spectra for each star.

The output combined spectra for all APOGEE objects are placed on the
same rest wavelength pixel scale, with a constant dispersion in 
$\log\lambda$, using 
$$\log \lambda_i = 4.179 + 6\times 10^{-6} i$$
with 8575 total pixels ($i=0$ to 8574), giving a rest wavelength range of 15100.8 to
16999.8 \AA. The dispersion was chosen to provide approximately 
3 pixels per resolution element, although the resolution varies
over the full wavelength range.

To do the combination, each rest frame (i.e., the Doppler shift is removed)
visit spectrum is sampled on this final wavelength scale using sinc interpolation.
Since the visit spectra are all dither combined, they are well sampled over the
entire range; in fact, they are significantly oversampled at the long wavelength
end.  The sinc interpolation takes this into account by using a chip-dependent
FWHM, conservatively adopted to be 5, 4.25, and 3.5 (dithered) pixels in the
red, green and blue chips, respectively. This effectively filters out noise
at higher spatial frequencies.

After a rough continuum normalization (using wide boxcar and median filters), the resampled spectra are then
combined using a weighted mean, with weights calculated on both a pixel-by-pixel and a spectrum-by-spectrum
basis (where the weight is the inverse square of the normalized error).
The resulting weighted average spectrum is then multiplied by the average continuum of the individual visit
spectra.  The determination of radial velocities and the spectral combination are done iteratively and is
described in more detail in Section \ref{subsubsec:relativervs}.

\begin{figure}[t!]
\begin{center}
\includegraphics[angle=0,scale=0.38]{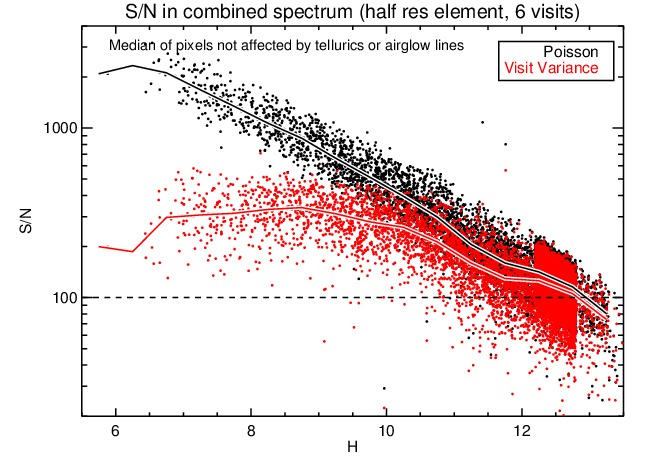}
\end{center}
\caption{The $S/N$ per half resolution element of pixels not affected by telluric absorption
  or sky lines derived using the pipeline noise model (black) and the variance
  in multiple measurements (red) for 9548 stars with 6 visits.  The lines show medians in bins
  of 1 mag.  The empirical S/N values are consistently low compared to the noise model values for
  high $S/N$s, which suggests that we are limited by systematic errors at the $\sim$0.5\% level for these stars.
}
\label{fig_apstar_snr}
\end{figure}

Figure \ref{fig_apstar_snr} shows the S/N values (per half resolution element) using
the pipeline noise model and the variance from multiple visits (for 9548 stars with
6 visits).  The empirical values are systematically low compared to the noise model estimates
in the high S/N regime (i.e., bright stars), which indicates
that we are limited by systematics (at the $\sim$0.5 \% level) for these stars.
However, we can easily achieve the S/N$\sim$100 required for the survey.

A combined LSF is created by taking a weighted average (using the same weights as above) of the individual visit LSF arrays on
the final apStar wavelength scale.  The model LSF (a sum of spatially-varying Gauss-Hermite functions;
see Section \ref{subsec:lsfcal}) is fitted to the empirical 2D LSF array and the coefficients of this
approximation are also saved.



\subsection{Output}

There is an output apStar-2MASSID.fits file for each unique object that contains the two combined spectra
(pixel-by-pixel and visit-by-visit weighting) as well as the individual visit spectra resampled on the apStar
wavelength scale.  The final, weighted LSF and the Gauss-Hermite coefficients are saved in apStarLSF-2MASSID.fits
files.  Summary information for all stars in a given field are saved in the apField-LOC4.fits files.

\section{Radial Velocity Determination}
\label{sec:rvs}

APOGEE radial velocities (RV) are derived at both the Visit and the Object stages.  The main steps are:

\begin{itemize}
\item As each visit is reduced, an RV estimate is determined by cross-correlating the visit spectrum against
a grid of synthetic spectra. This provides an ``estimated RV'' for the visit, which is stored in the apVisit files,
but not subsequently used since it was found to be unreliable sometimes.

\item ``Refined'' radial velocities for each visit are derived when the visit spectra are combined in the object stage.
This is done in three steps:
  \begin{enumerate}
  \item Relative radial velocities are determined using the combined spectrum as the spectral template. This is done
    iteratively because the relative RVs and the combined spectrum depend on each other.
  \item The absolute radial velocity of the combined spectrum is determined by cross-correlating it with a grid of
    synthetic spectra spanning a large range of stellar parameters.
  \item The relative radial velocities for each visit and the absolute velocity of the combined
    spectrum are then combined to produce absolute velocities for all visit spectra.
  \end{enumerate}
\end{itemize}

The latter scheme was employed because RVs derived from the combined spectrum (of the star itself) should be more
precise than RVs derived from a small set of synthetic spectra (although there can be issues for double-lined
spectroscopic binaries). This allows us to create a high-quality combined
spectrum without even knowing what type of object we are dealing with. However, the absolute RV is a critical
science product and the final combined spectrum must be on the rest wavelength scale so that it can be properly
compared to the large grid of synthetic spectra in the abundance pipeline (ASPCAP).  Therefore, the second step
in the ``refined'' RV determination is to derive the absolute radial velocity of the combined spectrum against a small grid
of synthetic spectra (the ``RV mini-grid'').

The various steps in the latter process are described in more detail below.

\begin{figure*}[ht!]
\begin{center}
\includegraphics[angle=0,scale=0.50]{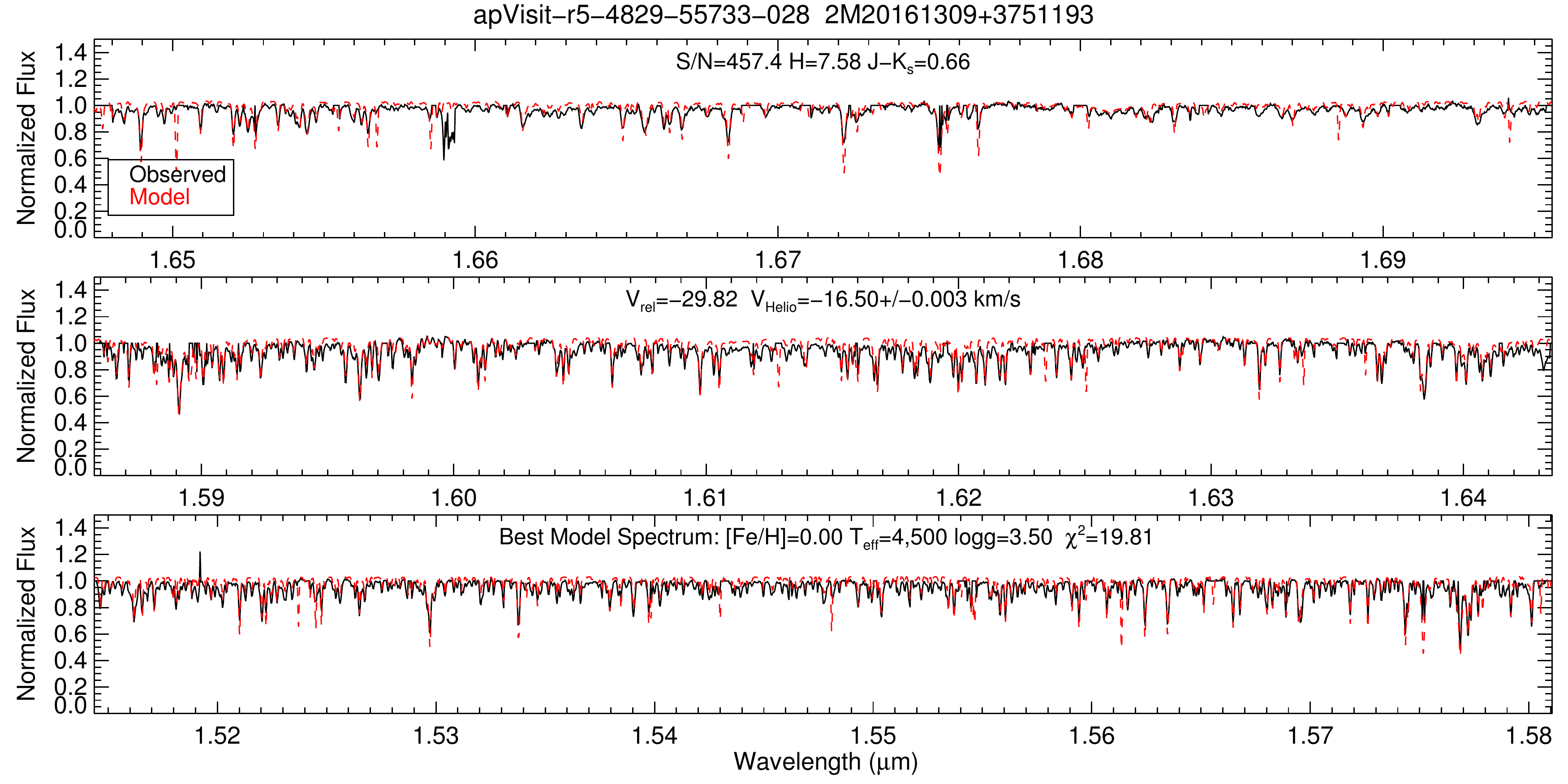}
\end{center}
\caption{Example APOGEE normalized spectrum (black) with best-fitting template spectrum (dashed red)
overplotted.  This is the ``estimated'' RV-fitting done at the individual visit level.}
\label{fig_exspec}
\end{figure*}

\subsection{Visit Stage Radial Velocities}
\label{subsec:visitrv}

At the end of the 1D Visit stage RVs are derived for all object spectra on the plate using the well-sampled
apVisit files.  This gives a first estimate for the RV (and therefore called ``estimated'' or EST RVs), but currently
these values are not used later on in the processing.  These were the only RVs available pre-DR10 and were used for
science papers such as \citet{Nidever12}.  We decided to keep this portion of the code in the pipeline so we could
compare RVs between the various methods.  For completeness we briefly describe how these estimates RVs are derived.

RVs are determined using a set of 96 synthetic template spectra (first generation ``RV mini-grid'') that
sparsely cover a large range in stellar parameters:
\begin{itemize}
\item 3,500 $<$ \teff $<$ 25,000 K
\item 2.0 $<$ \logg $<$ 5.0
\item $-$2.0 $<$ [Fe/H] $<$ 0.3
\end{itemize}
The ASS$\epsilon$T \citep{Koesterke08, Koesterke09, Koesterke12} spectral synthesis package and Kurucz model atmospheres
\citep{Castelli04, Meszaros12} were used for cool and warm stars (\teff $\leq$ 10,000 K) and Synspec \citep{Hubeny85, Hubeny11}
TLUSTY model atmospheres \citep{Hubeny88, Hubeny95, Lanz03, Lanz07} for hot stars (\teff $\geq$ 15,000 K).
Three steps are used to derive the RVs: (1) continuum normalization of observed and template spectra,
(2) cross-correlation of observed spectra against all template spectra, and (3) $\chi^2$-minimization using
the best-fitting template from step 2.

{\bf Normalization:}
Continuum normalization of the spectra is critical to obtain accurate RVs and especially important for
hot stars with very few, but wide, spectral features (i.e., Brackett lines).  The spectrum for each
of the three arrays is normalized separately.  First, bad pixels and pixels with bright airglow
lines are masked.  Next, the 95$^{\rm th}$ percentile is calculated in 40 spectral chunks (of $\sim$102 pixels each),
and then fit with a robust cubic polynomial.  The spectrum is normalized with this first estimate
of the continuum, and the binning and polynomial fitting are repeated to remove some
residual structure.  The final continuum is the product of the two polynomial fits.
The same procedure is used to normalize all of the synthetic template spectra.

{\bf Cross-correlation:}
Before the normalized observed spectrum can be cross-correlated with the template spectra it is
resampled (using cubic spline interpolation)
onto the logarithmic wavelength scale of the
template spectra.  The resampled observed spectrum is then cross-correlated with the template
spectrum, the best shift is found using the peak of the cross-correlation function, and
$\chi^2$ computed after shifting the template.  This procedure is repeated for all template spectra,
and the best-fitting template is chosen based on the lowest $\chi^2$ value.  To refine the
radial velocity determination, a Gaussian plus linear fit is performed on the peak of the cross-correlation function
of the best-fitting template.

{\bf $\chi^2$-minimization method:}
After the best-fitting template is found using cross-correlation, a second RV-determination technique
is used.  The observed spectrum is split up into 45 pieces ($\sim$273 pixels each)
and a separate
RV derived for each piece using $\chi^2$-minimization and the chosen template spectrum.  In this simple
forward-modeling technique, the only floating parameter is the Doppler shift.  At a given Doppler
shift the template spectrum is resampled onto the wavelength scale of the observed spectrum.
One advantage of the ``pieces'' technique is that an (internal) RV uncertainty can be calculated
directly from the multiple RV measurements of the pieces.
This technique works quite well for spectra of cool or metal-rich stars that have many narrow
lines, but less well for hot stars.

The final RV is chosen based on the calculated RV uncertainty of the two techniques.  A final
$\chi^2$ is calculated using the final, adopted Doppler shift and best-fitting template.  The barycentric
correction (see Section \ref{subsubsec:barycorr} below) is applied to convert the RVs to the
solar system barycenter.
An example APOGEE spectrum with the best-fitting template is shown in Figure \ref{fig_exspec}.

\subsection{Object Stage Radial Velocities}
\label{subsec:objectrv}

In the Object stage radial velocities are determined for all visits of a star together using
a common RV template.  The measurement of RVs and the spectral combination are performed iteratively
as described below.

\subsubsection{Preparing the Spectra}
The spectra are ``prepared'' for cross-correlation by:
\begin{itemize}
\item Pixel masking: Pixels marked as ``bad'' in the mask array or have sky lines in the sky array are masked out for the rest
of the RV determination.
\item Continuum normalization: Each of the three chip spectra are normalized separately. The chip spectrum is separated into
40 chunks (covering approximately 14 Angstroms each) and the 95$^{\rm th}$ percentile pixel value is calculated for each
chunk. A robust third-order polynomial is then fit to the chunk 95th percentile values. Finally, the spectrum is
normalized (divided) by the polynomial fit.  This is very similar to the Visit stage normalization method mentioned above.
\end{itemize}
This preparation process is performed on both the observed visit spectra as well as the RV template spectra
(observed combined or synthetic spectrum).

\subsubsection{Cross-Correlation}
All radial velocities are determined by cross-correlating a spectrum against a template spectrum. The spectra are on
the same logarithmic wavelength scale (see Section \ref{sec:apstar}) so that a Doppler shift is identical
to a constant shift in the x-dimension.  The spectra are ``prepared'' for cross-correlation by continuum normalization.
A Gaussian is fit to the peak of the cross-correlation function to more accurately determine the best spectral shift.
Finally, the shift and its uncertainty are converted to velocity units.

\subsubsection{Relative Radial Velocities}
\label{subsubsec:relativervs}
The relative radial velocities are determined by using the combined spectrum as the RV template. This is done iteratively,
first determining the relative RVs and then creating the combined spectrum using the relative RVs to shift the visit spectra
to a common (mean) velocity wavelength scale. For the first iteration, when no combined spectrum exists yet, the highest $S/N$
visit spectrum is used as the template. For all subsequent iterations the combined spectrum is used as the template. Each
iteration finds small shifts of the shifted and resampled visit spectra compared to the combined spectrum until the values
converge.

\subsubsection{Absolute Radial Velocities}
The combined spectrum after the relative RV step still has the mean RV of the star, which must be removed. The combined
spectrum is cross-correlated against each synthetic spectrum in the RV mini-grid. For each synthetic
spectrum the best RV and $\chi^2$ (of the shifted spectrum) are derived. The spectrum with the lowest $\chi^2$ is chosen
as the best-fitting spectrum and it's RV is used as the absolute RV of the combined spectrum.  Once the mean velocity
is determined the visit spectra are combined one last time with the mean velocity removed so that the final combined
spectrum is on the rest wavelength scale.

The second generation RV mini-grid is composed of 538 synthetic spectra that span a large range of stellar parameters:
\begin{itemize}
\item 2,700 $<$ \teff $<$ 30,000 K
\item 0.0 $<$ \logg $<$ 5.0
\item $-$2.5 $<$ [Fe/H] $<$ $+$0.5
\end{itemize}
However, the step sizes and ranges for \logg and [Fe/H] vary with effective temperature.
The ASS$\epsilon$T spectral synthesis package and Kurucz model atmospheres were used for cool and warm stars (3,500 $\leq$ \teff $\leq$ 14,000 K),
Synspec with TLUSTY model atmospheres for hot stars (\teff $\geq$ 15,000 K), and BT-Settl model spectra \citep{Allard11}
for very cool stars (2,700 $\leq$ \teff $\leq$ 3,300 K).  This new RV mini-grid covers
a larger range of parameter space and with finer sampling than the first generation grid.  In addition, a number of
spectra with high carbon and also high $\alpha$-elements are included to help serve as templates for carbon-rich and
oxygen-rich stars.  The  synthetic spectra have a resolution of 23,500 and are on the same logarithmically-spaced
wavelength scale as the APOGEE combined spectra.


We discovered that the BT-Settl spectra used for the coolest temperatures have a systematic,
temperature-dependent radial velocity offset on the order of 1 \kms \citep[][see Figure 7]{Cottaar14}.  The cause is not entirely clear but one
possibility is a wavelength shift of the molecular water lines in the BT-Settl linelist.  No RV corrections were applied
in DR11+12, but future releases will likely have the BT-Settl spectra removed from the mini-grid.

\subsubsection{Synthetic Radial Velocities}
After the best fitting template is determined, each individual visit spectrum is cross-correlated against this template
to derive what we call ``synthetic'' radial velocities.  We prefer the relative velocities derived (as discussed above) from
the cross-correlation of each visit with the combined spectrum, because this should be a better match that does not depend
on accuracy or completeness of the synthetic library.  However, this technique can perform poorly for faint stars because
the highest S/N visit spectrum (used as the template in the first iteration) is still quite noisy, and especially when
only a small number of the visit spectra are available so that the combined spectrum also has low S/N.  In APOGEE-2,
a significant number of faint halo stars and bulge RR Lyrae stars will be observed and, therefore, the RV algorithms will
need to be improved to accomodate these faint stars, likely by relying more heavily on synthetic spectra.

The synthetic RVs provide a check of the relative RVs for objects where there is a good library match.  The scatter
between the two types of RVs is stored in SYNTHSCATTER, and when this is larger than 1 \kmse, the
SUSPECT\_RV\_COMBINATION bit is set in the STARFLAG bitmask.  The cross-correlation functions from which the synthetic RVs
are derived are also useful for detecting double-lined spectroscopic binaries (SB2s) because the combined spectrum will be a
less helpful template in those cases.  No automatic SB2 identification algorithm is currently in use by the pipeline,
but the synthetic RV cross-correlation functions are saved in the apStar files and can be used for further inspection.

\subsubsection{Barycentric correction}
\label{subsubsec:barycorr}
Radial velocities in APOGEE are reported with respect to the center of mass of the Solar System - the barycenter. The
individual exposures are corrected for the relative motion of the Earth along the line-of-sight of the star during each
observation. This is called the ``barycentric correction'' and can be calculated very accurately (to \ms levels).
Our routines are partially based on those from \citet{McCarthy95}.  When these corrections are applied to the absolute
RVs from above we attain the RV with respect to the barycenter, or \vhelio for short.

\begin{figure}[t!]
\begin{center}
\includegraphics[angle=0,scale=0.45]{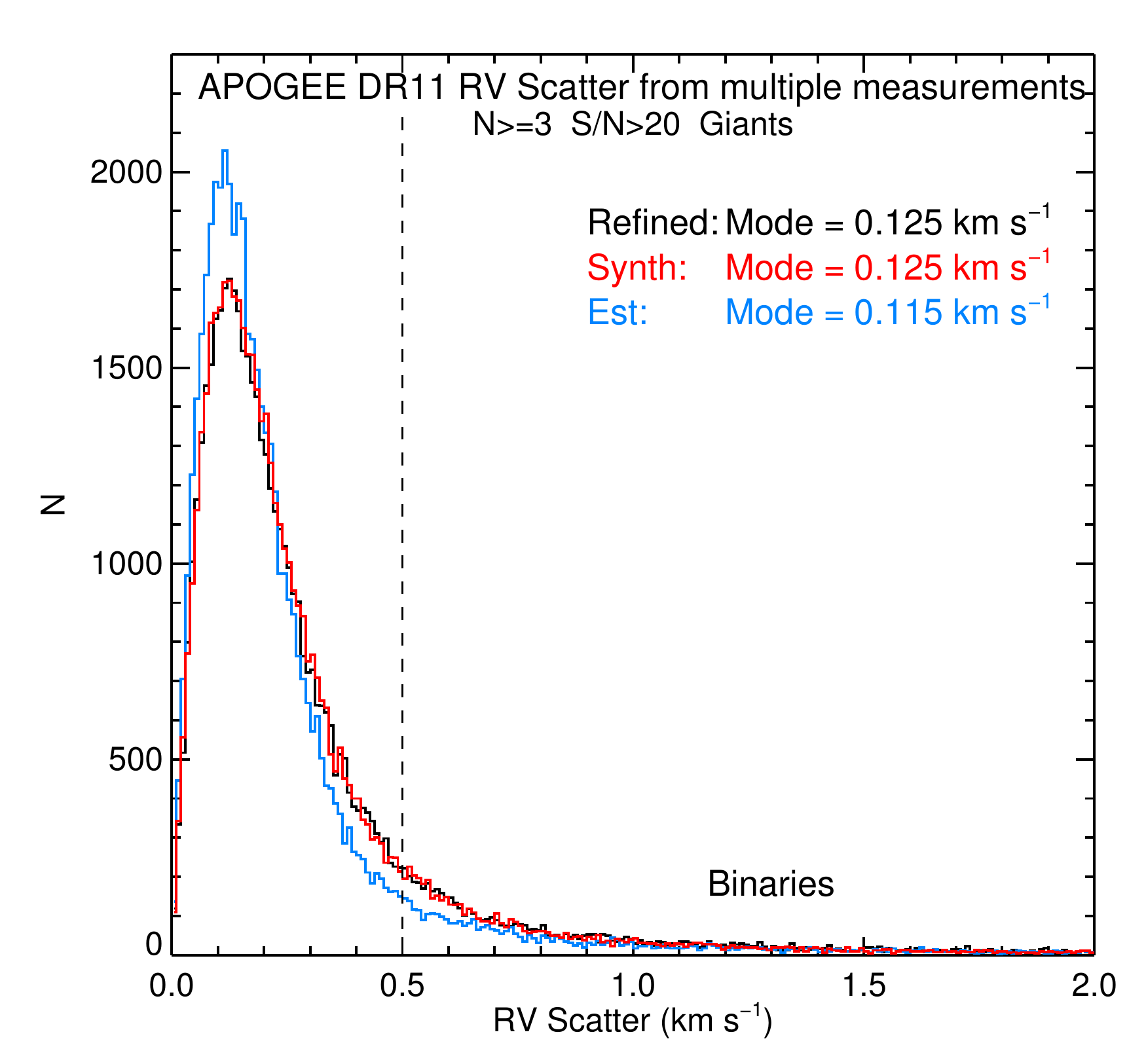}
\includegraphics[angle=0,scale=0.45]{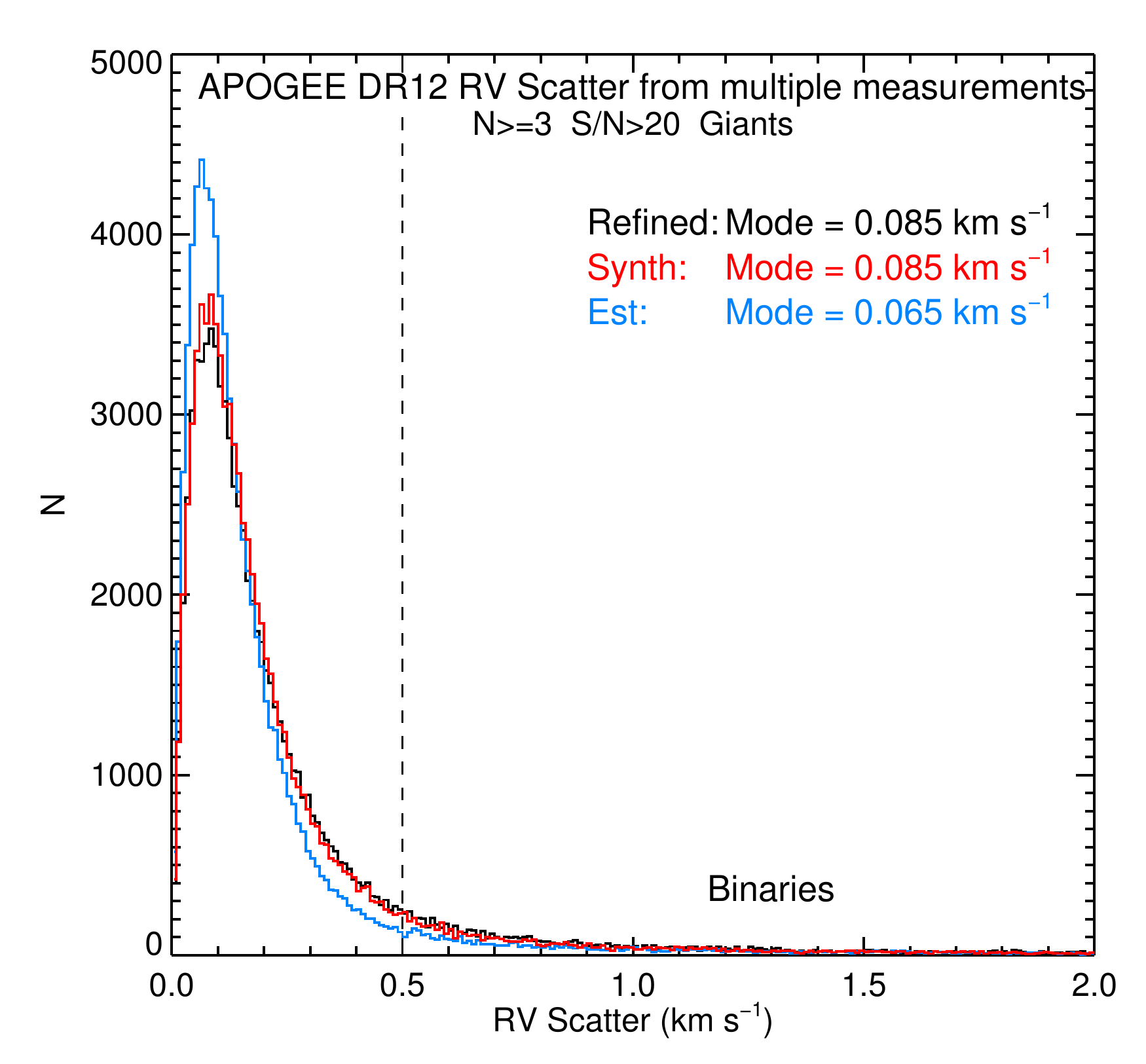}
\end{center}
\caption{Histograms of RV scatter in the three APOGEE RV measurements
(Refined: black, Synth: red, EST: blue) for giant stars with multiple visits
(N$\geq$3 and total $S/N\geq$20).  The distributions peak at $\sim$110 \ms for DRD11 (left)
  and at $\sim$80 \ms for DR12 (right).  The Refined and Synth methods have very
  similar distributions with slighter larger scatter values than EST in both DR11
  and DR12.  The RV scatter is a good internal measure of the the internal APOGEE
  RV precision and long-term stability.
Note that these values include real astrophysical variations due to binaries, which
likely explains the long tail.
}
\label{fig_rvscatter}
\end{figure}

\subsubsection{Absolute RV Zeropoint}
To ascertain the accuracy of the APOGEE velocity zero-point, we compared the APOGEE RVs to those of \citet{Nidever02}
and \citet{Chubak12}, which is on the Nidever velocity scale and accurate to $\sim$30 \ms.  For DR12, there are
41 unique stars in common between APOGEE and Nidever/Chubak (7 with Nidever et al.\ and 40 with Chubak et al.).  We find
\begin{align}
  \langle V_{\rm Nidever/Chubak} - V_{\rm DR12} \rangle = -0.355 \pm 0.033~\kms \nonumber
\end{align}
\noindent
with an rms scatter of 0.192 \kmse.
For DR11, there are only 15 stars in common and
\begin{align}
  \langle V_{\rm Nidever/Chubak} - V_{\rm DR11} \rangle = -0.615 \pm 0.089~\kms \nonumber
\end{align}
\noindent
with an rms scatter of 0.333 \kmse.
Therefore, the changes to the RV software from DR11 to DR12 improved both the zero-point (by $\sim$0.25 \kmse) and dispersion.
We find no clear trends with \teff, \logg or [Fe/H] and do not correct the RVs for any offsets.





\subsection{RV Uncertainties}
The RV uncertainty depends on the $S/N$, the resolution, and the information contained in the spectral lines themselves:
A spectrum with lots of deep and thin lines (such as in cool and metal-rich stars) will have a much more accurate RV
than a spectrum with few shallow and/or wide lines (such as for metal-poor or hot stars).  The cross-correlation RVs
estimate the velocity uncertainty from the uncertainty in the measurement of the cross-correlation peak, which is
partially set by the width of the peak.  This currently systematically underestimates the uncertainty in the RV
measurements and will be improved in the future.

\begin{figure*}[t!]
\begin{center}
  \includegraphics[angle=0,scale=0.31]{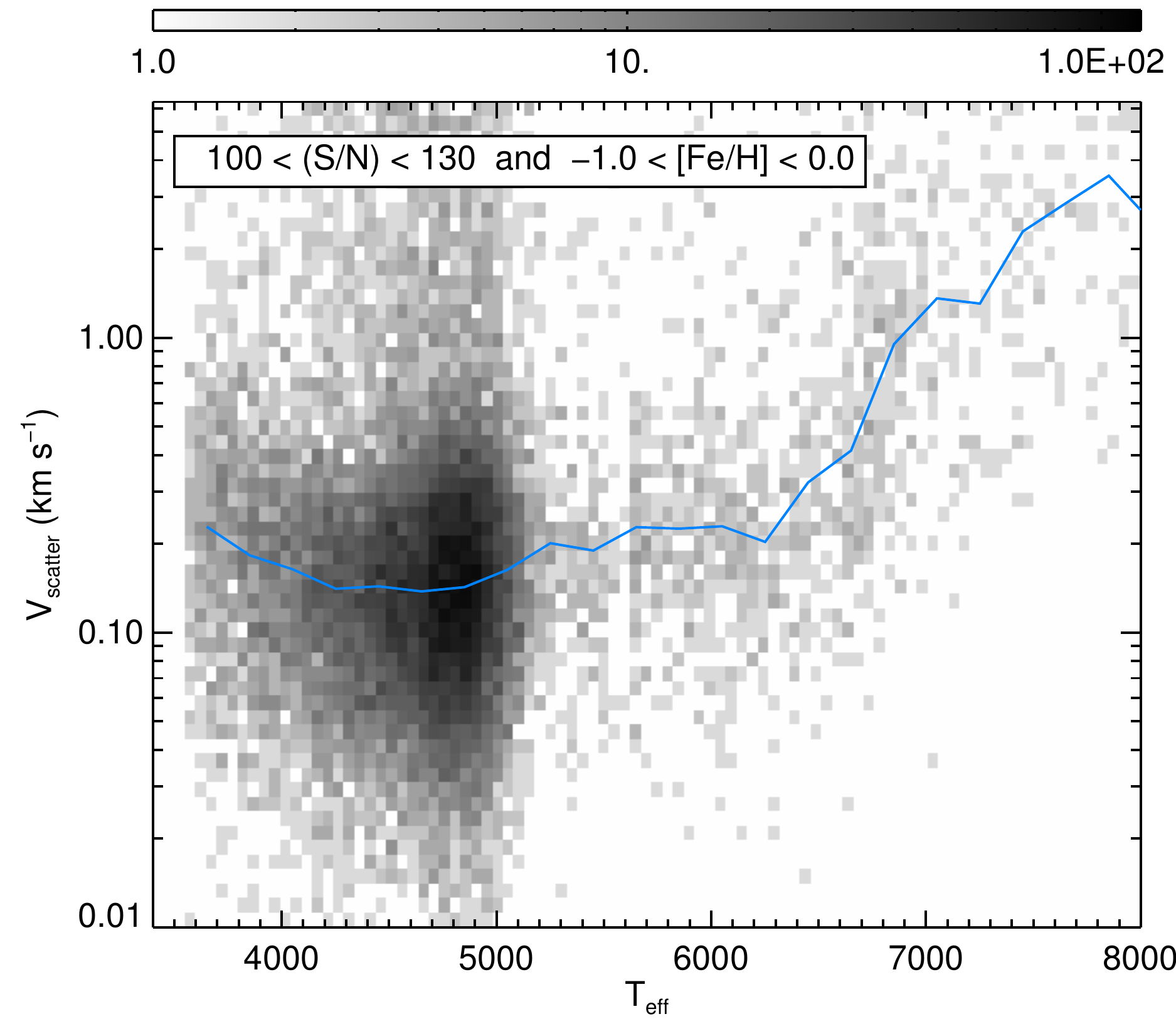}
  \includegraphics[angle=0,scale=0.31]{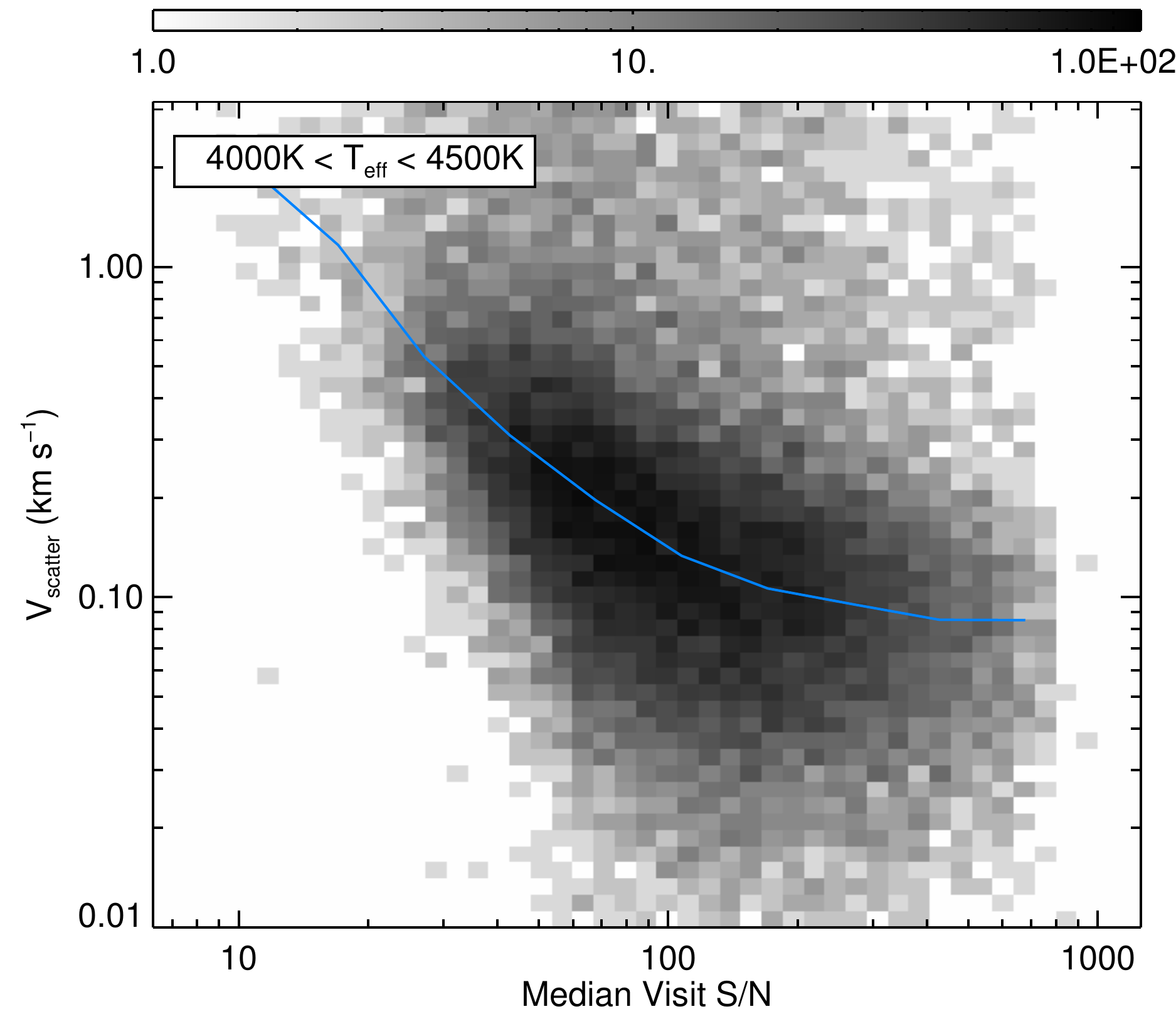}
  \includegraphics[angle=0,scale=0.31]{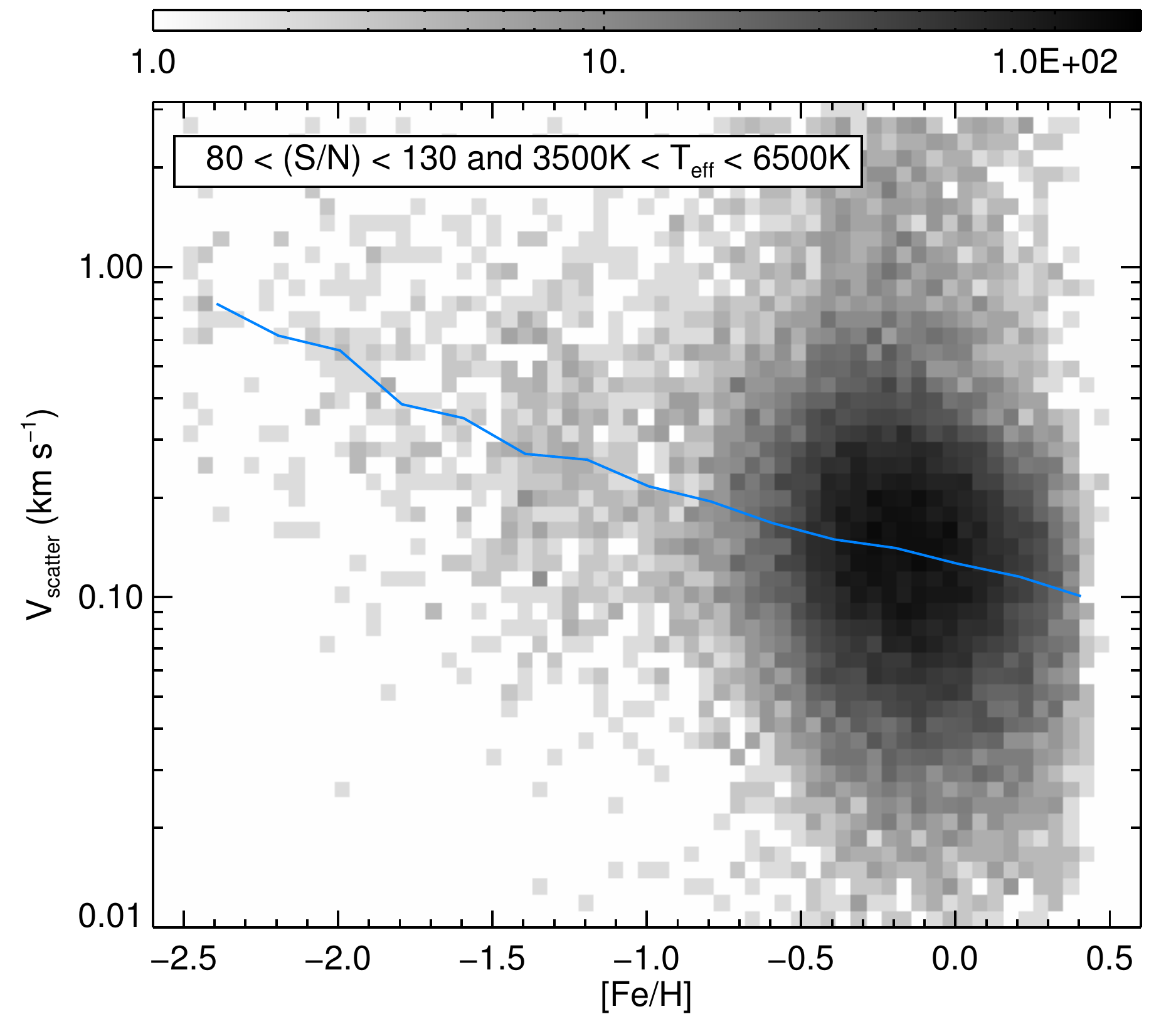}
\end{center}
\caption{The dependence of $V_{\rm scatter}$ on \teff (left), $S/N$ (middle) and [Fe/H] (right).
  Stars are selected to highlight the dependence in each panel.  The trends are as expected with
  the scatter increasing for higher \teffe, lower $S/N$, and metal-poor stars.
The blue line indicates median values in bins of the abscissa.
}
\label{fig_vscatter_relations}
\end{figure*}

A better estimate of the internal precision is the RV scatter for stars with multiple measurements.  For giants with a
total $S/N$$>$20 and three or more visits the distributions peak at $\sim$110 \ms for DR11 and $\sim$70 \ms for DR12
(Figure \ref{fig_rvscatter}).
The EST scatters are slightly smaller than for the Refined and Synth methods.  This is likely because the EST method uses
the visit spectra on their native and oversampled wavelength scale while both the Refined and Synth methods use the
resampled/downsampled visit spectra.  Future improvements on the RV software will likely use the visit spectra on their
native wavelength scale.  The RV scatter is higher for (1) dwarfs, because their lines are broadened by rotation and higher
surface gravities, and, (2) especially for hotter stars with broader lines.  Figure \ref{fig_vscatter_relations} shows the
dependence of $V_{\rm scatter}$ on \teffe, $S/N$, and metallicity.

\begin{figure}[t]
\begin{center}
\includegraphics[angle=0,scale=0.38]{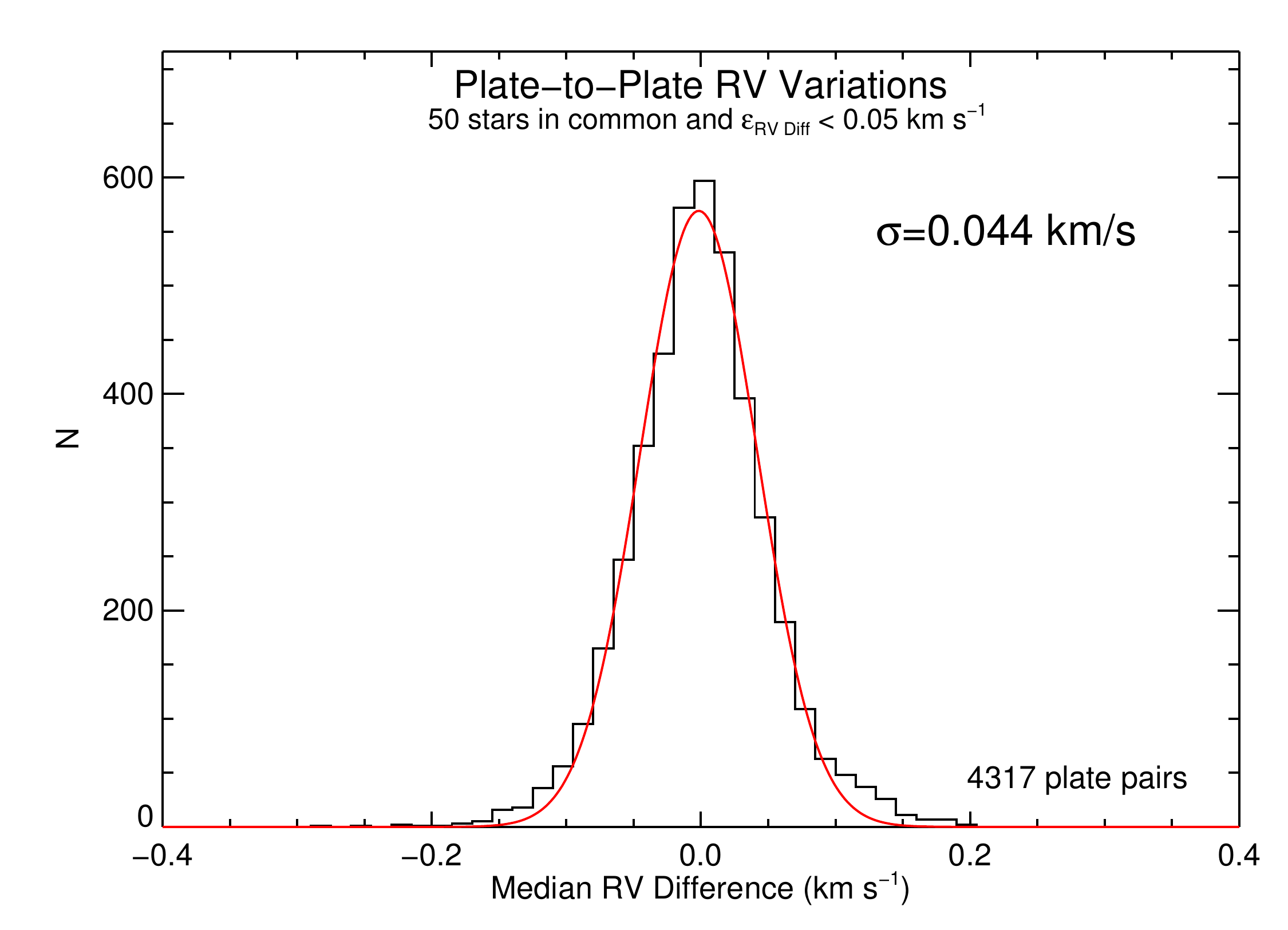}
\end{center}
\caption{The distribution of plate-to-plate velocity differences for 4317 plate pairs with 50 or more stars
  in common and an uncertainty in the velocity difference of less than 0.05 \kmse.  The rms scatter is 0.044 \kms
  indicating that the RVs are very stable.}
\label{fig_rvoffset_plates}
\end{figure}

Another estimate of the internal accuracy and long term stability of the APOGEE RVs are median plate-to-plate RV differences
using stars in common.  Figure \ref{fig_rvoffset_plates} shows the histogram of median RV differences of 4317 plate pairs
with more than 50 stars in common and RV difference uncertainties less than 0.05 \kmse.  The distribution is centered
around zero and has an rms scatter of 0.044 \kmse.  This indicates that the APOGEE instrument and the RVs have been
very stable over the three years of survey operations.


\subsection{Binarity}
The pipeline currently has no flag for binarity because the RV uncertainties are underestimated.  In addition, giants have
significant astrophysical RV scatter (``jitter'') especially at the tip of the RGB \citep[e.g.,][]{Hekker08} that makes
binary classification more complicated.  Work is ongoing to develop a reliable binary classification scheme.  However,
in the meantime large RV scatter may point to binarity.  The first catalog of stellar and substellar companions to APOGEE stars
will released in the near future (Troup et al. 2015, in preparation).

\section{Data Access}
\label{sec:access}

All of the data are accessible through the SDSS-III Science Archive Server 
(SAS), which provides web access to the entire set of data products, ranging 
from the raw data cubes to the reduced spectra (including intermediate data products). These are organized
in a directory structure and files that are described by the SDSS-III
datamodel, which can be viewed at \path{http://data.sdss3.org/datamodel}.
The SAS can be accessed at \path{http://data.sdss3.org/sas}, with the top
level for APOGEE raw data at
\path{http://data.sdss3.org/sas/dr12/apogee/spectro/data}, and the top
level for APOGEE reduced data products at
\path{http://data.sdss3.org/sas/dr12/apogee/spectro/redux}.
Abbreviated descriptions of the APOGEE project, instrument, software, and data products can be found on the
SDSS-III DR12 web site (\path{http://www.sdss.org/DR12}) and summary file descriptions can be found in the
SDSS-III data model.
 
The main products of the reduction pipeline are the reduced visit spectra,
which are stored in apVisit files, and the reduced combined spectra, which
are stored in apStar files. A webapp interface to download these for
individual targets and bulk target lists is available at
\path{http://data.sdss3.org}. Otherwise, users can navigate their way through
the SAS directory structure to find individual files and intermediate
data products.

Note that most of the image/spectra data products are multi-extension FITS
files, where the contents of the extensions are described in the datamodel.
For IDL users, the SDSS-III apogeereduce software product contains a 
routine, APLOAD.PRO, that reads all of the extensions with a single command
and stores the results in an IDL data structure. 

Parameters extracted from the spectra, along with object information, are
stored in summary FITS table files. The allVisit file contains information
about individual visits, while the allStar file contains information from
the combined spectra. These tables are also loaded in the the Catalog
Archive Server (CAS), a database with a web interface at 
\path{http://skyserver.sdss3.org}.

\label{sec:datamodel}

The official APOGEE data model -- defining the directory structure, filename convention,
file formats, and header keywords -- is available from
\path{http://data.sdss3.org/datamodel/files/APOGEE_ROOT/}.

\section{Summary}
\label{sec:summary}

We have described the automated APOGEE data reduction pipeline. The basic steps are:
\begin{enumerate}
\item Collapsing the 3D up-the-ramp data cube to a 2D image and removing cosmic rays.
\item Extraction of the 300 spectra from the 2D image and rough flux calibration.
\item Wavelength calibration, sky and telluric correction, combination of dither pairs, and absolute flux calibration.
\item Combination of visit spectra and radial velocity determination.
\end{enumerate}

Some of the non-standard features of the pipeline are:
\begin{itemize}
\item Cosmic ray rejection:  Having up-the-ramp data gives us the ability to detect and remove most
  of the cosmic rays in our data.
\item Dither combination: The APOGEE spectra are slightly undersampled and, therefore, APOGEE observations
  are taken as pairs with a half pixel spectral dither between them.  These spectra are ``dither-combined''
  with a sinc-interlace interpolation to create a single well-sampled spectrum.
\item Model telluric correction:  We fit fairly simple telluric absorption models to hot star spectra across the plate
  to derive ``scaling'' values for each telluric species.  Two-dimensional polynomial fits are then performed
  of the variations of these scalings across the plate and subsequently used to derive a model telluric
  absorption spectrum for each observed science spectrum.
\item Iterative relative RV determination:  We use the combined observed spectrum of each star as it's own
  RV template to derive precise relative RVs.  This is performed in an iterative fashion since the relative RVs
  and the combined spectrum depend on each other.
\end{itemize}

Future improvements include better sky subtraction, persistence correction, and updates to the RV routines for
low-$S/N$ spectra.

\acknowledgements


D.L.N. was supported by a McLaughlin Fellowship at the University of Michigan.
Sz.M. has been supported by the J{\'a}nos Bolyai Research
Scholarship of the Hungarian Academy of Sciences.
C.A.P. is thankful to the Spanish MINECO for support through grant AYA2014-56359-P.
We thank the anonymous referee for useful comments that improved the manuscript.
Funding for SDSS-III has been provided by the Alfred P. Sloan
Foundation, the Participating Institutions, the National Science
Foundation, and the U.S. Department of Energy Office of Science. The
SDSS-III web site is http://www.sdss3.org/.
SDSS-III is managed by the Astrophysical Research Consortium for the
Participating Institutions of the SDSS-III Collaboration including the
University of Arizona, the Brazilian Participation Group, Brookhaven
National Laboratory, Carnegie Mellon University, University of
Florida, the French Participation Group, the German Participation
Group, Harvard University, the Instituto de Astrofisica de Canarias,
the Michigan State/Notre Dame/JINA Participation Group, Johns Hopkins
University, Lawrence Berkeley National Laboratory, Max Planck
Institute for Astrophysics, Max Planck Institute for Extraterrestrial
Physics, New Mexico State University, New York University, Ohio State
University, Pennsylvania State University, University of Portsmouth,
Princeton University, the Spanish Participation Group, University of
Tokyo, University of Utah, Vanderbilt University, University of
Virginia, University of Washington, and Yale University.


%

\end{document}